\documentclass[12pt]{iopart}

\usepackage{harvard} 
\usepackage{here}

\usepackage{graphicx}
\begin{document}

\title[Review of short-range gravity experiments in the LHC era]{Review of short-range gravity experiments in the LHC era}

\author{Jiro Murata and Saki Tanaka}

\address{
Department of Physics, Rikkyo University, 3-34-1 Nishi-Ikebukuro, Tokyo 171-8501, Japan}
\ead{jiro@rikkyo.ac.jp}
\begin{abstract}
This document briefly reviews recent short-range gravity experiments that were performed at below laboratory scales to test the Newtonian inverse square law of gravity. To compare sensitivities of these measurements, estimates using the conventional Yukawa parametrization are introduced. Since these experiments were triggered by the prediction of the large extra-dimension model, experiments performed at different length scales are compared with this prediction. In this paper, a direct comparison between laboratory-scale experiments and the LHC results is presented for the first time. A laboratory experiment is shown to determine the best limit at $M_D > 4.6 \;\rm{TeV}$ and $\lambda<23 \;\mu \rm{m}$. In addition, new analysis results are described for atomic systems used as gravitational microlaboratories. 
\end{abstract}

\submitto{\CQG}

\section{Introduction}
\label{section-intro}
It has been more than 300 years since Newton published \textit{Principia}, introducing the Newtonian universal gravitational law \cite{principia}.
Unlike the case of Coulomb's law, at the time the gravitational law was introduced,
there was no direct experimental data confirming the validity of the inverse square law.
Even now, the law remains the least precisely tested of all fundamental physical laws.
In fact, Newtonian gravitational constant has been determined only to a precision of  $10^{-4}$, while, for example, the fine structure constant $\alpha_F$ is known to a precision of $10^{-10}$.
Here, Newtonian gravity $F_N$ is defined by
\begin{equation}
F_N=G_\infty \frac{Mm}{r^2},
\end{equation}
where $G_\infty$ is the gravitational constant for two point masses, $M$ and $m$, that are separated by distance $r$.
In this paper, $G_\infty$ refers to the ideal gravitational constant that has no $r$ dependence.
In addition to the value of $G_\infty$, the distance dependence of the gravitational force $F\propto 1/r^2$ also remains untested to good precision, especially at small and large scales compared with the scale of the Earth-Moon system. As shown below, the gravitational inverse square law has been tested to a precision of $10^{-10}$ only near the scale of the Earth-Moon system. This should be compared with the precision of Coulomb's law, which is known to have $q<10^{-16}$ in the form of $F\propto 1/r^{2+q}$  \cite{PhysRevLett.26.721,PhysRevA.33.759}.

Although it is obviously inadequate to assume that Newton's gravitational law applies far beyond the region that has been experimentally tested, we still tend to do so without question. 
For an example, for the Planck mass
\begin{equation}
M_{pl}=\sqrt{\frac{\hbar c}{G}}=1.22\times 10^{16} \; {\rm TeV/c^2}
\end{equation}
we assume that the gravitational constant remains constant at $G=G_{\infty}$ down to a very small scale of the Planck length, $L_{pl}=10^{-35}$ m.
This bold estimate extrapolates the inverse square law over a scale of more than $10^{30}$ from the experimentally tested region.
Compared with the other three interactions, our experience with gravity is very limited.
Indeed, no one has succeeded in observing a gravitational phenomenon below 10 $\mu$m.
Therefore, we can say that even the existence of gravity has not yet been confirmed at microscopic scales.

In recent decades, there have been several key predictions and experimental claims regarding non-Newtonian gravity.
(1) Fujii's dilaton model prediction in 1971, (2) Long's claim in 1976 of evidence for a violation of the inverse square law, (3) Fischbach's claim in 1986 of a composition-dependent gravity known as the ``fifth force," and (4) the 1998 prediction of non-Newtonian gravity based on a large extra-dimension model.

In 1971, Fujii proposed a possible non-Newtonian gravity that violates the inverse square law by assuming a dilaton-meditated new interaction with a range from 10 m to 1 km, or below 1 cm \cite{fujii}.
Fujii predicted that the gravitational potential should be
\begin{equation}
V_{Fujii}(r)=-G_\infty \frac{Mm}{r} ( 1+\frac{1}{3}e^{-r/\lambda} ) = G_{Fujii}(r) \frac{Mm}{r},
\label{fujii-func}
\end{equation}
where $\lambda$ is understood to be an interaction range for the new dilaton field. 
For $\lambda \gg r_N$, $G_N=G_{Fujii}(r_N)=\frac{4}{3}G_\infty$ is obtained using an $r$-dependent gravitational constant $G_{Fujii}(r)$.
In this paper, we denote $G_N$ as the experimental Newtonian gravitational constant measured at a distance corresponding to a laboratory scale of $r_N\equiv 0.1$ m.
Fujii assumed that $G_\infty = \frac{3}{4}G_N$ may differ from the measured value $G_N$, which is obtained by assuming the functional form for the Newtonian gravitational potential,
\begin{equation}
V_N(r_N)=-G_N \frac{Mm}{r_N}.
\end{equation}
Fujii's Yukawa-type expression can be generalized as
\begin{equation}
V_{Yukawa}(r)=-G_\infty \frac{Mm}{r} ( 1 + \alpha e^{- r/\lambda} ).
\label{Yukawa}
\end{equation}
The Yukawa range $\lambda$ can also be written as $\lambda=\frac{\hbar}{m_b c}$ using the mass $m_b$ of the exchanged boson.
If $m_b \rightarrow 0$, (\ref{Yukawa}) reverts to the Newtonian inverse square law.
For example, if there is an additional new type of massive graviton, the total gravitational potential should take the form of the Yukawa potential in (\ref{Yukawa}).
The Yukawa potential can also be expressed in terms of a distance-dependent gravitational ``constant"
\begin{equation}
G_{Yukawa}(r)=G_\infty ( 1 + \alpha e^{- r/\lambda} ).
\label{Yukawa-G}
\end{equation}
Then, the relation between $G_N$ and $G_\infty$ is
\begin{equation}
G_N=G_{Yukawa}(r_N)
\rightarrow \left\{
\begin{array}{lll}
&G_\infty (1+\alpha)& (r_N \ll \lambda)\\
&G_\infty & (\lambda \ll r_N)
\end{array}.
\right.
\label{Yukawa-GN-Ging}
\end{equation}
For Fujii's specific prediction, $G_N=\frac{4}{3}G_\infty$ is obtained when $\alpha=1/3$ and $\lambda \gg r_N$.
We can assume $G_\infty=G_N$ only when $\lambda \ll r_N$.
For other cases, we cannot, in general, use the measured value $G_N$ as $G_\infty$ in (\ref{Yukawa}).

\begin{figure}[t]
 \begin{center}
  \includegraphics[width=100mm]{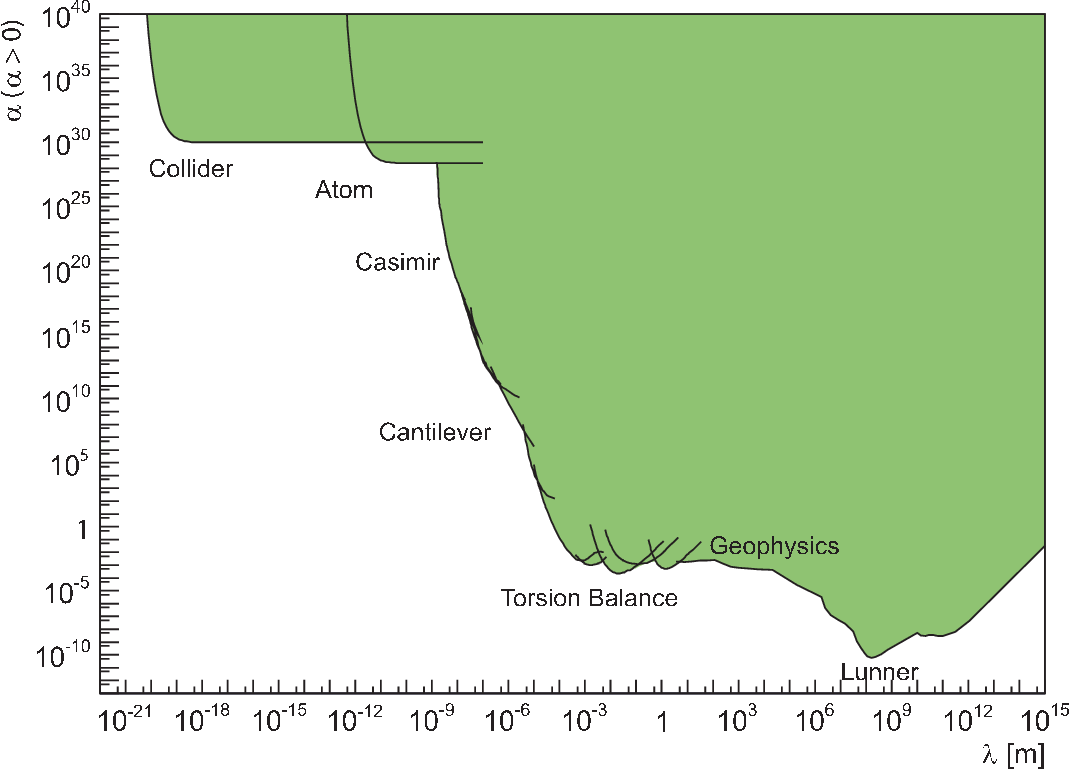}
 \end{center}
 \caption{Experimental constraints on the parameters $\alpha$ (coupling strength) and $\lambda$ (range) of Yukawa interaction for $\alpha\;(>0)$. Shaded area indicates excluded area at 95\% confidence level. Constraint curves for over km scales are taken from \cite{fischbach1999search,Adelberger:2009zz}. See Section \ref{sec-experiment} for short-range tests at below laboratory scale.}
 \label{alpha-lambda-fullscale}
\end{figure}

Triggered by Fujii's proposal, a number of modern experiments were performed at geophysical ($\sim$ km) and laboratory ($\sim$ m) scales.
The experimental constraints on the Yukawa parameters are shown in Figure \ref{alpha-lambda-fullscale}.
This so-called $\alpha-\lambda$ plot was first introduced by Talmadge as a model-independent expression of experimental constraints \cite{PhysRevLett.61.1159}.

In 1976, Long claimed that he had found evidence for a distance dependence in $G$ of the form $G_{Long}(r)=G_\infty (1+0.002 \ln{[r/\rm{cm}]})$ at a cm scale \cite{Nature260}.
Many laboratory Cavendish-type experiments tried to confirm his result; however, all attempts failed to confirm any violation of the Newtonian inverse square law.
Then, in 1986, Fischbach claimed that there must be composition dependence in $G$; this was based on the reanalysis of the classic Et\"{o}v\"{o}sh experiment data \cite{Eotvos1922}. This argument is known as the ``fifth force," which can be expressed as
\begin{equation}
G_{5}(r)=G_\infty ( 1 + \tilde{\alpha} q_1 q_2 e^{- r/\lambda} ),
\label{5th}
\end{equation}
where $q_i$ is a generalized point charge for particles $i=1, 2$, divided by their masses $m_i$, which are normalized by the hydrogen mass $m_H$ \cite{PhysRevLett.56.3}.
For example, if the additional Yukawa interaction couples to baryon numbers,
$q_i = B_i / (m_i / m_H)$ is obtained for the $i$-th composition using its atomic mass $m_i$ and baryon number $B_i = A_i$ (atomic number).
The existence of this new composition-dependent gravitational force, i.e., nonzero $\tilde{\alpha}$, would lead to a violation of universality of free fall at short range.
Many experiments tried to find evidence of a composition-dependent $G$ over various length scales.
Although this fifth force proposals essentially introduces a violation of the universality of $G$, experimental constraints on $\tilde{\alpha}$ can be set without directly testing for composition dependence using different material combinations; instead, all that is required is merely the test for distance dependence.
Therefore, all modern experimental tests on the inverse square law can also test this fifth force proposal. 
Of course, such tests cannot provide evidence that $\tilde{\alpha}$ exists; they can only set upper limits.
Following these attempts to find deviations from the Newtonian gravitational inverse square law, since the beginning of 1990's, we have come to believe that the inverse square law is confirmed to a precision of $10^{-4}$ at cm scale.
Very detailed descriptions are summarized in the textbooks written by Fischbach and Talmadge \cite{fischbach1999search} and by Franklin \cite{franklin1993fifth}.

\begin{figure}[h]
 \begin{center}
  \includegraphics[width=100mm]{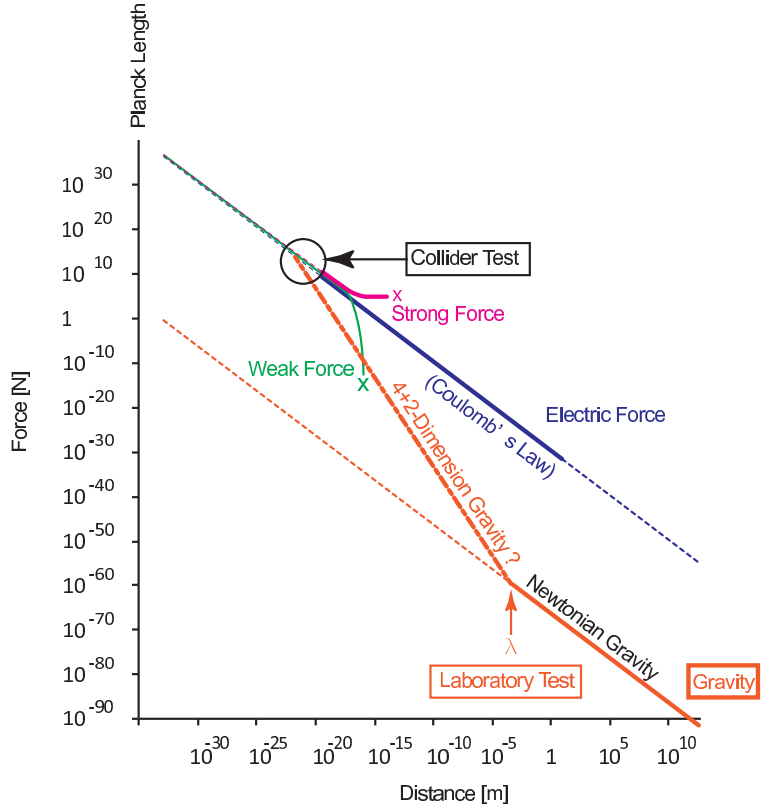}
 \end{center}
 \caption{Illustration of four fundamental force strengths between two quarks or electrons. Electromagnetic, strong, weak and gravitational force strengths are shown, as functions of their distance \cite{HE_News.32.2014}.}
 \label{force-strength}
\end{figure}

Then in 1998, a striking model of large extra-dimensions was proposed by Arkani-Hamed, Dimopoulos, and Dvali (ADD model). This model predicts a violation of the gravitational inverse square law at approximately 0.1 mm, provided that there are two additional ``large" spatial dimensions outside our four-dimensional space-time; these extra-dimensions were invoked to naturally resolve the so-called hierarchy problem
\cite{Arkani.Hamed1998263,PhysRevD.59.086004}.
Figure \ref{force-strength} illustrates the force strength $F$ of the four fundamental interactions as functions of distance $r$ between two quarks or two electrons.
The linear slopes represent the inverse square law $F\propto 1/r^2$.
Although Figure \ref{force-strength} is not based on rigorous calculation, we can grasp a rough estimate of the relation between the fundamental interactions.
Note that the gravitational inverse square law has been precisely tested only at above laboratory scale, where the line is solid; the region wherein the gravitational force has not been observed is shown as a dashed line.
In contrast, the electric force law has only been tested at below laboratory scale.
The other two interactions, the strong and weak interactions, tend, by a renormalization calculation, to have the same order of force strength as the electric force at around $10^{-20}$ m.
From Figure \ref{force-strength}, the hierarchy problem can be visually interpreted  as the big gap between gravity and the three other interactions at below collider scales of $10^{-19}$ m, which corresponds to a de Broglie wave length $\lambda=h/p$ at TeV energies.

In the ADD model, the modified potential
\begin{equation}
V_{4+2}(r)=-G_{4+2}\frac{Mm}{r^3}
\end{equation}
is expected at $r<0.1 \; \rm{mm}$ by assuming a higher dimensional Planck mass of $M_D \sim 1 \;\rm{TeV}$.
Here, $G_{4+2}$ is a new (4+2)-dimensional gravitational constant.
More generally, the ADD potential can be expressed as
\begin{equation}
V_{ADD}(r)= \left\{
\begin{array}{cc}
-G_{4+n}\frac{Mm}{r^{1+n}} &(r<\lambda) \\
-G_\infty \frac{Mm}{r} &(r>\lambda)
\end{array}
\right.,
\label{ADD-potential}
\end{equation}
using the size of the large extra dimensions $\lambda$.
Here, $G_{4+n}=G_\infty \lambda^n$ is obtained from a connecting condition at $r=\lambda$.
In this model, the modification of the inverse square law is expected at a scale smaller than $r_N$, i.e., $\lambda \ll r_N$; therefore, we can use $G_N$ as $G_\infty$ in the expressions.
(\ref{ADD-potential}) can be understood by assuming an isotropic three-dimensional space viewed from a large scale and an isotropic (3+n)-dimensional space at a small scale by Gauss's law.
The expression of these in a single smooth functional form is more convenient,
\begin{equation}
V_{power}(r)=-G_\infty \frac{Mm}{r} \left[1+\left(\frac{\lambda}{r}\right)^n \right].
\label{power}
\end{equation}
Obviously, for  $r \ll \lambda$, (\ref{power}) cannot be approximated by the Yukawa form in (\ref{Yukawa}).
For example, at very short scales, modification factors from the Newtonian potential are
$V_{power}/V_N= \left[1+(\lambda/r)^n \right]
\rightarrow \infty$; however,
$V_{Yukawa}/V_N= \left[1+\alpha e^{-r/\lambda} \right]
\rightarrow 1+\alpha$,  
at $r/\lambda \rightarrow 0$.
Since all results from experimental tests after Fujii's proposal have been expressed in the Yukawa parametrization, the latest generation precision experiments at below mm scales, which are searches for a signal of the large extra-dimension, have also tried to constrain the parametrization of ($\alpha$, $\lambda$) in (\ref{Yukawa}).
Therefore, the direct comparisons of the experimental sensitivities of the ADD model using the Yukawa potential between different experiments, which were performed at very different separation distances, are confusing.
Instead, in this paper, we propose to use the power-law parametrization in tests of the ADD model.
In the following sections, we describe how to compare model sensitivities based on both the Yukawa and power-law parametrizations.

\section{General formalism of the experimental tests}
\label{sec-formalism}
Let us first consider how to treat non-Newtonian gravity in a general way. 
In principle, we cannot perform a direct measurement of the absolute potential $V(r)$.
Instead, a differential of the potential, i.e., the force, is usually the quantity to be measured. 
A modified gravitational force can be written as
\begin{equation}
F(r)=G_\infty \frac{Mm}{r^2} ( 1 + a(r) ),
\label{force-general}
\end{equation}
where the additional term $a(r)$ represents the $r$-dependence of the gravitational constant. 
Then, a generalized gravitational constant
\begin{equation}
G(r)=G_\infty ( 1 + a(r) )
\label{G-general}
\end{equation}
is obtained.
Experimental confirmation of $a(r)\neq 0$ implies a nonconstant $G(r)$.
If the distance-varying $G(r)$ is experimentally measured as a function of $r$ at many distance points, then the functional form of $a(r)$ can be determined.
At present, in the search for the first evidence of a nonzero $a(r)$, we can test (\ref{G-general}) just by comparing $G(r)$ at two different measuring distances, $r_{near}$ and $r_{far}$.
If we define new experimental parameters
\begin{equation}
\gamma\equiv\frac{G(r_{near})}{G(r_{far})}, \; \delta \equiv \gamma-1,
\label{gamma}
\end{equation}
deviation from $\gamma=1$ or $\delta=0$ implies the violation of the inverse square law.
In general, if an experimental result for $\gamma$ is obtained, we assume a model for $a(r)$ with unknown parameters, and then, evaluate them, for example, $(\alpha,\lambda)$ in the Yukawa parametrization or $(n, \lambda)$ in the power-law parametrization.

\subsection{Measurement of $\gamma$}
Here we consider the methods to experimentally measure $\gamma$. One way is to measure an absolute value of $G(r)$ and then compare it with results from other experiments performed at other length scales (absolute measurements). The other way is to perform an experiment that includes measurements at different distances (relative measurements). 

\subsubsection{Absolute measurements.}
The simplest way to determine $\gamma$ is to measure the gravitational constant $G(r_{exp})$ at $r_{exp}$ as $r_{near}$ and then compare it with results from other experiments, such as $G_N=G(r_N)$ at $r_N$ as $r_{far}$. 
Then, we can form
\begin{equation}
\gamma=\frac{G(r_{exp})}{G_N}.
\end{equation}
The experimental precision for $\gamma$ is limited by that of $G_N$ at around $10^{-4}$; therefore, this treatment can be used only at or above approximately 0.1 \% precision in $\gamma$.
Needless to say, this approach cannot be applied for a length scale at $r_{exp} \sim r_N$.

To determine an absolute value for $G(r_{exp})$, we need to know the absolute values of the measured force, masses, and distances to be used in
\begin{equation}
G(r_{exp})=F_{exp}(r_{exp}) \frac{r_{exp}^2}{Mm}.
\end{equation}
To avoid the uncertainties in these experimental parameters, we should try relative measurements where many of the uncertainties cancel one another.
Based on the current experimental precision of the $\alpha-\lambda$ plot shown in Figure \ref{alpha-lambda-fullscale}, this absolute measurement may be useful at very short scales below 0.1 mm.

\subsubsection{Relative measurements.}
A smart way to avoid the uncertainties in absolute measurements is to perform a set of measurements that include at least two different distances in the same experiment. 
Then, $\gamma$ can be obtained directly from 
\begin{equation}
\gamma=\frac{G(r_{near})}{G(r_{far})}=\frac{F_{exp}(r_{near})}{F_{exp}(r_{far})} \left(\frac{r_{near}}{r_{far}}\right)^2  \left(\frac{M_{far}m_{far}}{M_{near}m_{near}}\right)^2,
\end{equation}
using only the ratios of experimental parameters. 
Here, $M_{near(far)}, m_{near(far)}$ are the masses used in the near and far configurations.
For example, we can avoid the demanding absolute determinations of the masses.
For absolute measurements, if the measurements are performed using torsion balances, cantilevers, or other elastic body devices, their spring constants need to be determined to obtain absolute values for the forces.
However, in relative measurements, the spring constants are the same for the near and far settings; therefore, they do not have to be precisely determined to obtain $\gamma$.
Such relative measurements are a powerful way to achieve a precise determination below the currently known precision of $G_N$. 
Furthermore, for $r \sim r_N$, a relative measurement is the only way to test for the existence of $a(r)$.

\subsubsection{Null measurements.}
Many relative measurements are, in fact, performed as so-called null experiments.
These experiments are designed to keep  
\begin{equation}
\left(\frac{r_{near}}{r_{far}}\right)^2  \left(\frac{M_{far}m_{far}}{M_{near}m_{near}}\right)^2 = 1,
\end{equation}
under the expectation that $F_{exp}(r_{near})=F_{exp}(r_{far})$ for Newtonian gravity.
Therefore, any imbalance between $F_{Newton}(r_{near})$ and $F_{Newton}(r_{far})$ implies $\gamma \neq 1$.
Such null measurements are sometimes useful for improving the precision of force measurements.
This is because a measuring dynamic range can be set as small as possible to obtain a large measuring gain (magnification).
Application of the null measurement itself does not improve the precision; however, it can reduce the measuring dynamic range, which can help improve the reading resolution.
Nevertheless, if the precision is dominated by random mechanical movements, a reduced dynamic range does not improve precision.

\begin{figure}[h]
 \begin{center}
  \includegraphics[width=100mm]{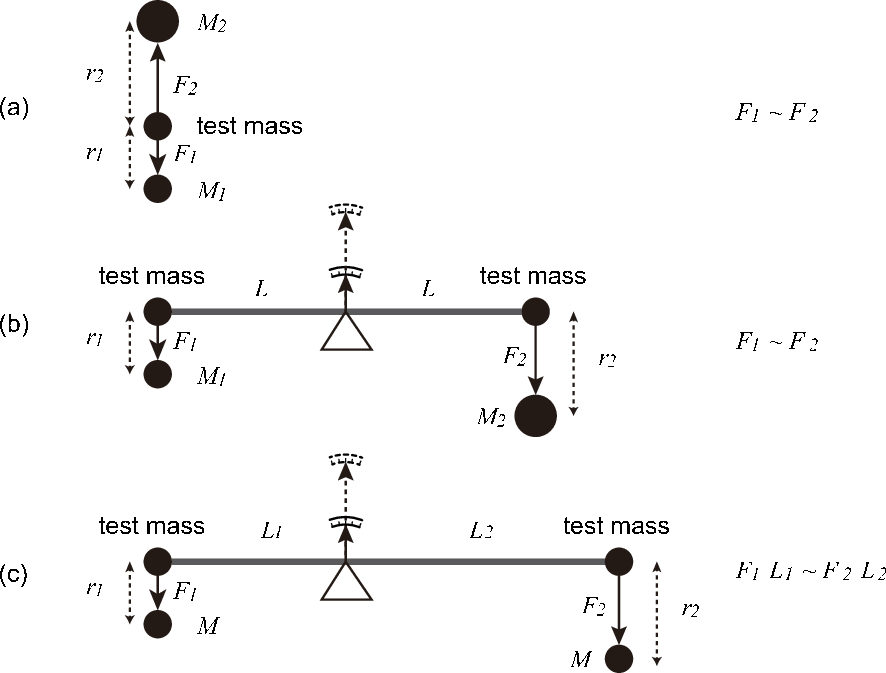}
 \end{center}
 \caption{Illustration of the null measurements: (a) two different source masses at different distances, (b) same as (a), but using a symmetric balance bar, (c) same as (b) but using asymmetric balance bar.}
 \label{Null-Figure}
\end{figure}

In Figure \ref{Null-Figure}, three examples of null configurations are shown using balances. In these experiments, torques from the two different source masses, which are located at different distances from the nearest test mass, are designed to be balanced. If we ignore cross forces between the test and source masses, which are set on opposite ends of a balance bar for simplicity, the null conditions can be written as $F_1=F_2$ for (a) and (b), and $F_1 L_1=F_2 L_2$ for (c).

The null condition means that such measurements are ``static," without large dynamic movements due to gravity.
Therefore, we can avoid nonlinearity of the spring constant in the measuring device.
Such relative measurements can avoid uncertainties in the absolute determination of spring constants; however, experimental determination should be carefully performed by including nonlinearities, i.e., possible deviations from Hooke's law.
Such null measurements can avoid nonlinearity problems in calibration between measured displacements and applied force strengths.

\subsection{Model parametrizations}
Next, we consider how a measured value of $\gamma$ can be interpreted in terms of model parameters. 
We consider two types of parametrization: the Yukawa parametrization defined in (\ref{Yukawa}) and the power-law parametrization defined in (\ref{power}). 
For simplicity, we start the treatment assuming point masses. 
Then, finite size corrections are discussed.

\subsubsection{Yukawa parametrization.}
\begin{figure}[t]
 \begin{center}
  \includegraphics[width=100mm]{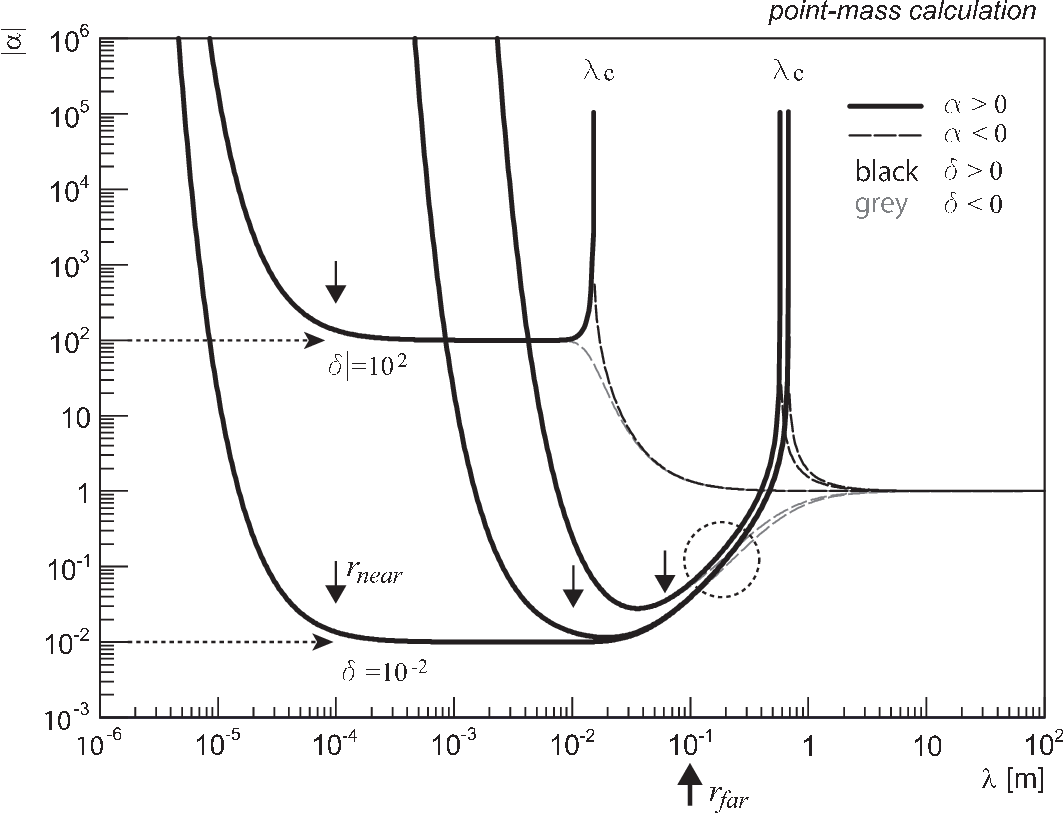}
 \end{center}
 \caption{Typical examples of $\alpha-\lambda$ function in Yukawa parametrization for point mass calculations. The minimum points are given at $\alpha=\delta$, except for the case of $r_{near}\sim r_{far}$. For all the cases,  $r_{far}=0.1 \; \rm{m}$  is fixed. The left ``hinge" points represent the positions at  $\lambda=r_{near}$. 
Three cases of $r_{near}=5 \;\rm{cm}, 1 \;\rm{cm}, 0.1 \;\rm{mm}$ for $\delta=10^{-2}$, and a case of $r_{near}=0.1 \;\rm{mm}$ for $\delta=10^2$ are shown. Black (gray) lines indicate for cases of $\delta>0$ ($\delta<0$). If $\alpha$ is positive (negative), they are indicated as solid (dashed) lines. In cases for $\delta>0$, there are critical points $\lambda_c$, where $\alpha \rightarrow \infty$. The dotted circle indicates ``weak rising" at $\lambda>r_{far}$, defined in 
equation (\ref{alpha-Yukawa-mod})}.
 \label{Yukawa-model}
\end{figure}

Since the Yukawa potential can represent boson exchange forces, many proposed new physics models can be expressed in this parametrization. 
See \cite{Adelberger:2009zz} and \cite{Long199923} for the proposed models.
The large extra-dimension model can also be parametrized using the Yukawa form for the lowest order diagram wherein $r\sim \lambda$.
First, we show how experimental constraints can be determined.
In (\ref{force-general}), $a(r)$ is expressed as
\begin{equation}
a(r)=\alpha ( 1+\frac{r}{\lambda} ) e^{- r/\lambda};
\label{ar-Yukawa}
\end{equation}
therefore, 
\begin{equation}
\gamma=\frac{1+a(r_{near})}{1+a(r_{far})}=
\frac{1+\alpha ( 1+\frac{r_{near}}{\lambda} ) \; e^{- r_{near}/\lambda} }
{1+\alpha ( 1+\frac{r_{far}}{\lambda} ) \; e^{- r_{far}/\lambda} }.
\label{gamma-Yukawa}
\end{equation}
By solving this equation, we can obtain an expression for $\alpha$ as a function of $\lambda$
\begin{equation}
\alpha=\frac{\delta}
{
(1+\frac{r_{near}}{\lambda})\; e^{-r_{near}/\lambda}
-
(\delta+1) (1+\frac{r_{far}}{\lambda}) \; e^{-r_{far}/\lambda}
},
\label{alpha-Yukawa}
\end{equation}
with $\delta=\gamma-1$.
Then, if we can ignore uncertainties on $r$, experimental constraint on $\alpha$ can be set as
\begin{equation}
\alpha(\delta=\delta_{min}) < \alpha < \alpha(\delta=\delta_{max}),
\end{equation}
using the experimental value of $\delta$, including its error.
Then (\ref{alpha-Yukawa}) can be re-written as
\begin{equation}
\alpha=\frac{\delta \; e^{r_{near}/\lambda}}
{
(1+\frac{r_{near}}{\lambda})
-
(\delta+1) (1+\frac{r_{far}}{\lambda})\; e^{-(r_{far}-r_{near})/\lambda}
}.
\label{alpha-Yukawa-mod}
\end{equation}
The numerator can be understood as a rapid falling of $\alpha$ at $\lambda < r_{near}$, and the denominator can be understood as a relatively weak rising at $\lambda > r_{far}$ in the $\alpha-\lambda$ plot.
This function has a minimum at around $\lambda \sim r_{near}, r_{far}$.
Typical examples of (\ref{alpha-Yukawa-mod}) are shown in Figure \ref{Yukawa-model}.

As shown in Figure \ref{Yukawa-model}, when $r_{near} \ll r_{far}$, the minimum occurs around $\alpha \sim \delta$.
Therefore, the minimum in $\alpha$ roughly represents the precision of $\delta$ in the measurement.
For absolute measurements using $G_N$ as $G(r_{far})$, the corresponding experimental constraints should be limited to $\lambda<r_N$, because $r_N$ is not usually well defined.
Therefore, in absolute measurements, there should not be a weak rising in $\alpha$ at large $\lambda$.
This implies that a single exponential form provides a good approximation for absolute measurements at small measuring distances $r_{exp} \ll \lambda$,
\begin{equation}
\alpha=\frac{\delta }{1+\frac{r_{exp}}{\lambda}} e^{r_{exp}/\lambda} 
\sim \delta \; e^{r_{exp}/\lambda} \;\;\;
 (r_{exp} \ll \lambda).
\label{alpha-Yukawa-abs}
\end{equation}
All experiments at below $\mu$m scales can be treated in this simple form, provided that the object is small enough to ignore the finite size effect described in Section \ref{finite-size}.

\subsubsection{Power-law parametrization.}
\begin{figure}[t]
 \begin{center}
  \includegraphics[width=100mm]{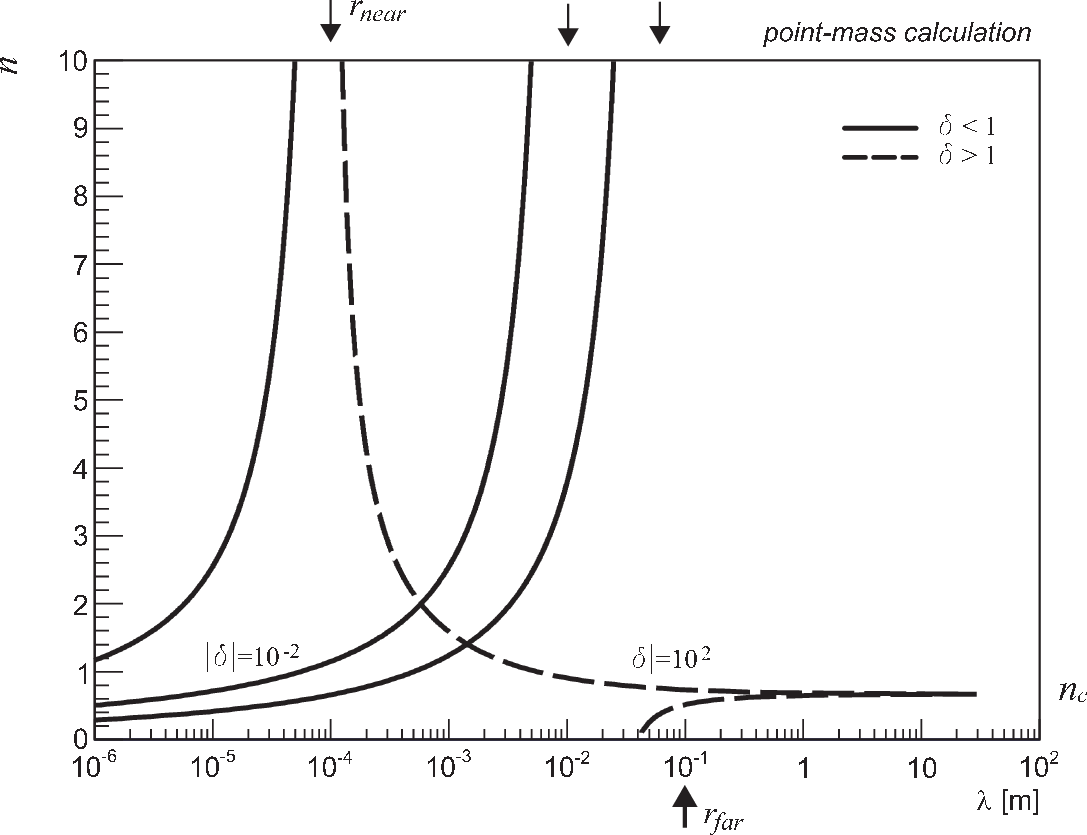}
 \end{center}
 \caption{
Typical examples of $n-\lambda$ function in the power law parametrization for point mass calculations. 
For all the cases,  $r_{far}=0.1 \; \rm{m}$  is fixed. 
The left ``kink" points represent the positions at  $\lambda=d_{near}$.
Three cases of $r_{near}=5 \;\rm{cm}, 1 \;\rm{cm}, 0.1 \;\rm{mm}$ for $\delta=10^{-2}$, and a case of $r_{near}=0.1 \;\rm{mm}$ for $\delta=10^2$ are shown. Solid lines indicate precision measurements $\delta<1$, and dashed line indicates a rough measurement $\delta>1$. In cases for $\delta>1$, there are critical points $n_c$, where $\lambda \rightarrow \infty$.
 }
 \label{power-model}
\end{figure}

In addition to the large extra-dimension model, other models also predict non-Newtonian gravity in the power-law form; these are introduced in \cite{doi:10.1142/S0217751X02013356}.
For the power-law parametrization, $a(r)$ in (\ref{force-general}) is expressed as
\begin{equation}
a(r)=(1+n)\left( \frac{\lambda}{r} \right)^n.
\end{equation}
In general, the exponent $n$ does not have to be an integer.
In the ADD model, $n$ represents the number of large extra-dimensions, which is usually treated as an integer $n \leq 6$. 
Just as for the Yukawa parametrization, the corresponding $\gamma$ can be expressed as
\begin{equation}
\gamma=\frac{1+a(r_{near})}{1+a(r_{far})}=
\frac{1+(1+n)\left( \frac{\lambda}{r_{near}} \right)^n }
{1+(1+n)\left( \frac{\lambda}{r_{far}} \right)^n },
\label{gamma-power}
\end{equation}
which can be rewritten as
\begin{equation}
\lambda=
\left(
\frac{
\delta/(1+n)
}
{
r_{near}^{-n}-(\delta+1)r_{far}^{-n}
}
\right)^{1/n}
=
\left(
\frac{
\delta/(1+n)
}
{1
-(\delta+1)(r_{near}/r_{far})^{n}
}
\right)^{1/n}
r_{near}.
\label{n-power}
\end{equation}
Typical examples of this function are shown in Figure \ref{power-model}.
For comparison, the parameter set in Figure \ref{power-model} is the same as that used in Figure \ref{Yukawa-model}.
Here, only attracting cases ($\alpha>0$ in the Yukawa parametrization) is treated in  the power-law parametrization.

As shown in Figure \ref{power-model},  at large $n$, $\lambda$ is limited to $r_{near}$.
Note that the slopes are opposite for large $\delta\; (>1)$ and small $\delta\; (<1)$.
Therefore, for small $n$, a precision measurement with small $\delta$ can set constraint at very small $\lambda$ region relative to the experimental scale $r_{near}$; however,  a rough measurement with large $\delta$ can set constraint only at very large $\lambda$ regions.
As we will see in Section \ref{add-discussion}, this is why, at small $n$, high-energy collider and short-range gravity experiments meet each other at the same 0.1 mm scale , in the context of the large extra-dimension.

\subsubsection{Finite size effects.}
\label{finite-size}
In real laboratory experiments, all objects have finite sizes. Since we are not assuming the inverse square law, we cannot assume that forces between finite objects are the same as those between point masses, even for spherical objects.
The modified potential must be volume integrated over each object; however, $G(r)$ can be regarded as almost constant at $G_N$, if $\lambda$ is very small compared with the center-to-center distance $r$ or to the surface-to-surface separation (gap) $d$ of the objects.
Therefore, finite size effects need to be considered for $\lambda$ at only above around the experimental scale.
This effect is most significant for thick parallel plates with a small separation gap $d$.

\begin{figure}[H]
 \begin{center}
  \includegraphics[width=100mm]{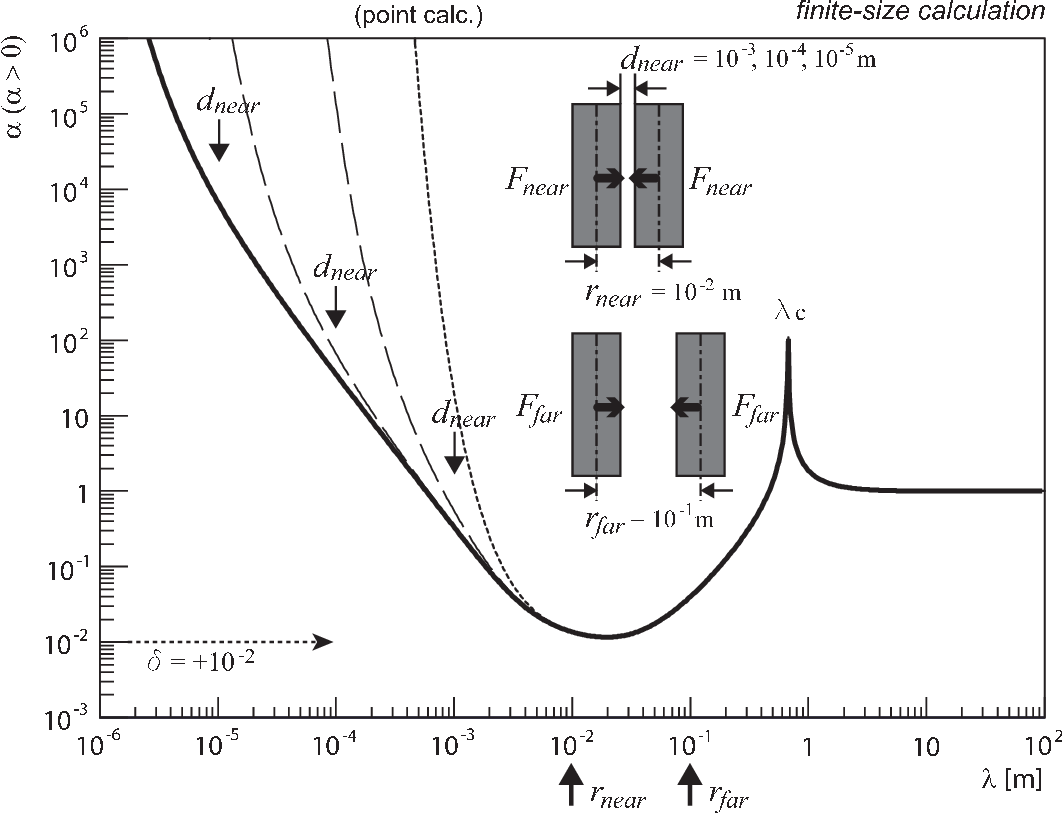}
 \end{center}
 \caption{$\alpha-\lambda$ plot in Yukawa parametrization for infinite sized parallel plates. $r_{near}=1 \; \rm{cm}$, $r_{far}=10 \; \rm{cm}$ and $\delta=10^{-2}$ are fixed for all the three cases, which separation gaps are $d_{near}=10^{-3}, 10^{-4},10^{-5} \; \rm {m}$, and $d_{far}=d_{near}+r_{far}-r_{near}$. Dotted line shows corresponding point-mass calculation as a reference.}
 \label{Yukawa-model-plate}
\end{figure}

Here, two parallel plates with infinite sizes are considered.
The force density (for a unit area) between infinite plates with thicknesses $l$ and placed at separation gap of $d$ can be calculated by
\begin{equation}
\mathcal{F}=\int_0^l dx \int_0^l dX \; 2\pi \rho^2 \; G_{\infty} (1+a(r)),
\end{equation}
where $\rho$ is mass volume density of the two plates, while $x$ and $X$ the vertical coordinates penetrating each plate in the outward direction. Then, the distance between small volume elements at $x$ and $X$ is $d+x+X$.
Just as for point mass calculations, measurements at two different distances
$d=d_{near}$ and $d_{far}$ are required.
Then, the force ratio at $d_{near}$ and $d_{far}$ can be expressed as
\begin{eqnarray}
\frac{\mathcal{F}(d_{near})}{\mathcal{F}(d_{far})}&=&
\frac{2\pi G_{\infty} \rho^2 l^2 + 2\pi G_{\infty} \rho^2 \int_0^l \int_0^l  a(d_{near}+x+X) dx dX}{2\pi G_{\infty} \rho^2 l^2 + 2\pi G_{\infty} \rho^2 \int_0^l \int_0^l a(d_{far}+x+X) dx dX} \nonumber \\
&=&\frac{
1+\frac{1}{l^2}\int_0^l \int_0^l  a(d_{near}+x+X) dx dX
}{
1+\frac{1}{l^2}\int_0^l \int_0^l  a(d_{far}+x+X) dx dX
}.
\label{finite-general}
\end{eqnarray}
For the Yukawa parametrization, this equation can be solved as
\begin{equation}
\alpha=\frac{
\delta_f l^2
}{
\int \int d_{near} - (\delta_f+1) \int \int d_{far}
},
\end{equation}
where
$\delta_f=\mathcal{F}(d_{near})/\mathcal{F}(d_{far})-1$, and
\begin{eqnarray}
\int \int d 
&\equiv & \int_0^l \int_0^l (1+\frac{d+x+X}{\lambda}) e^{-\frac{d+x+X}{\lambda}} dx dX \nonumber \\
&=& e^{-\frac{d}{\lambda}} \lambda^2 (1-e^{-\frac{l}{\lambda}})
\left[
(1+\frac{d}{\lambda})(1-e^{-\frac{l}{\lambda}})
+2\left(1-(1+\frac{l}{\lambda})e^{-\frac{l}{\lambda}}\right)
\right].
\end{eqnarray}
Finally, $\alpha$ can be obtained by
\begin{equation}
\alpha=\frac{\delta_f}{V_f}
e^{\frac{d_{near}}{\lambda}} \frac{l^2/\lambda^2}{1-e^{-l/\lambda}},
\label{finite-Yukawa}
\end{equation}
where the volume correction term $V_f$ is defined as
\begin{eqnarray}
V_{f}&=&
\left[
(1+\frac{d_{near}}{\lambda})(1-e^{-\frac{l}{\lambda}})+2\left(1-(1+\frac{l}{\lambda})e^{-\frac{l}{\lambda}}\right)
\right] 
-(\delta_f+1) \nonumber \\
&\times &
\left[
(1+\frac{d_{far}}{\lambda})(1-e^{-\frac{l}{\lambda}})
+2\left(1-(1+\frac{l}{\lambda})e^{-\frac{l}{\lambda}}\right)
\right]
e^{-\frac{d_{far}-d_{near}}{\lambda}}.
\end{eqnarray}
The first exponential factor in (\ref{finite-Yukawa}) represents a rapid falling at $\lambda < d_{near}$.
The finite size effect is most significant at the ``kink" point at $\lambda \sim d_{near}$, which is shifted down from the ``hinge" point at $\lambda \sim r_{near}$ in the point-mass calculation.
These experiments involve very short distance measurements between point masses separated by $r \sim d_{near}$.
Then, a constraint similar to that from point-mass calculation at $r_{near}=d_{near}$ is yielded.
Typical examples are shown in Figure \ref{Yukawa-model-plate}.
Clear kink positions at $\lambda \sim d_{near}$ can be seen.
Here, the $r_{near}$ and $r_{far}$ positions are defined as center-to-center distances, as shown in Figure \ref{Yukawa-model-plate}.

\vspace{1cm}
\begin{figure}[H]
 \begin{center}
  \includegraphics[width=100mm]{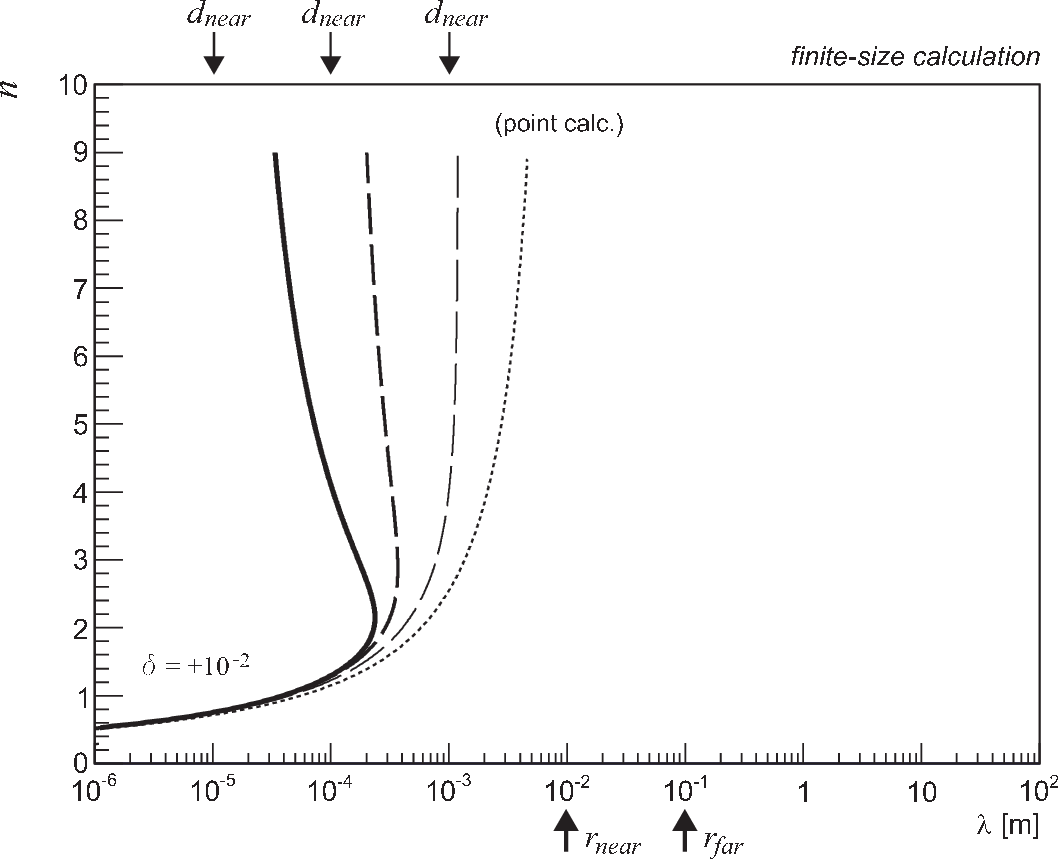}
 \end{center}
 \caption{$n-\lambda$ plot in power-law parametrization for infinite sized parallel plates. $r_{near}=1 \; \rm{cm}$, $r_{far}=10 \; \rm{cm}$ and $\delta=10^{-2}$ are fixed for all the three cases, which separation gaps are $d_{near}=10^{-3}, 10^{-4},10^{-5} \; \rm {m}$, and $d_{far}=d_{near}+r_{far}-r_{near}$. Dotted line shows corresponding point-mass calculation as a reference.}
 \label{power-model-plate}
\end{figure}

For the power-law parametrization, we have
\begin{equation}
\lambda=\left(
\frac{
\delta_f l^2
}{
\int \int d'_{near} - (\delta_f+1) \int \int d'_{far}
}
\right)^{1/n}
,
\end{equation}
where
\begin{equation}
\int \int d' 
\equiv \int_0^l \int_0^l \frac{1+n}{(d+x+X)^n} dx dX
\end{equation}
\renewcommand{\arraystretch}{1.5}
\begin{equation}
= \left\{
\begin{array}{cc}
\frac{1+n}{(n-1)(n-2)}
\left[
(d+2l)^{2-n}-2(d+l)^{2-n}+d^{2-n}
\right] & (n\neq 1,2) \\
2\left[(d+2l)\ln{(d+2l)}-2(d+l)\ln{(d+l)}+d \ln{ d } \right]& (n=1) \\
3\left[ -\ln (d+2l)+2\ln(d+l)-\ln d \right] & (n=2). 
\end{array}
\right.
\end{equation}
Typical examples are shown Figure \ref{power-model-plate}.
The figure shows the shifting of $\lambda$ values at large $n$ from $r_{near}$ to $d_{near}$. 

Figures \ref{Yukawa-model-plate} and \ref{power-model-plate} show that finite size effects are significant for the configuration $d_{near}\ll r_{near}$.
Contributions from closest volume elements can be regarded as point-mass contributions mixed in the measurement.
Any finite size effect can be regarded as a superposition of point-mass calculations.
In general, a numerical calculation of the volume integration should be performed to obtain the appropriate constraints on the model parameters for a specific experimental configuration; however, characteristics of the finite size effects can be understood in the present  parallel-plate calculations. 

\section{Experimental constraints}
\label{sec-experiment}

Based on the model parametrization analyses in Section \ref{sec-formalism}, analyzed results from existing experiments can be interpreted as constraints on the parameter space.

\subsection{Constraints on the Yukawa parametrization}
\begin{figure}[t]
 \begin{center}
  \includegraphics[width=100mm]{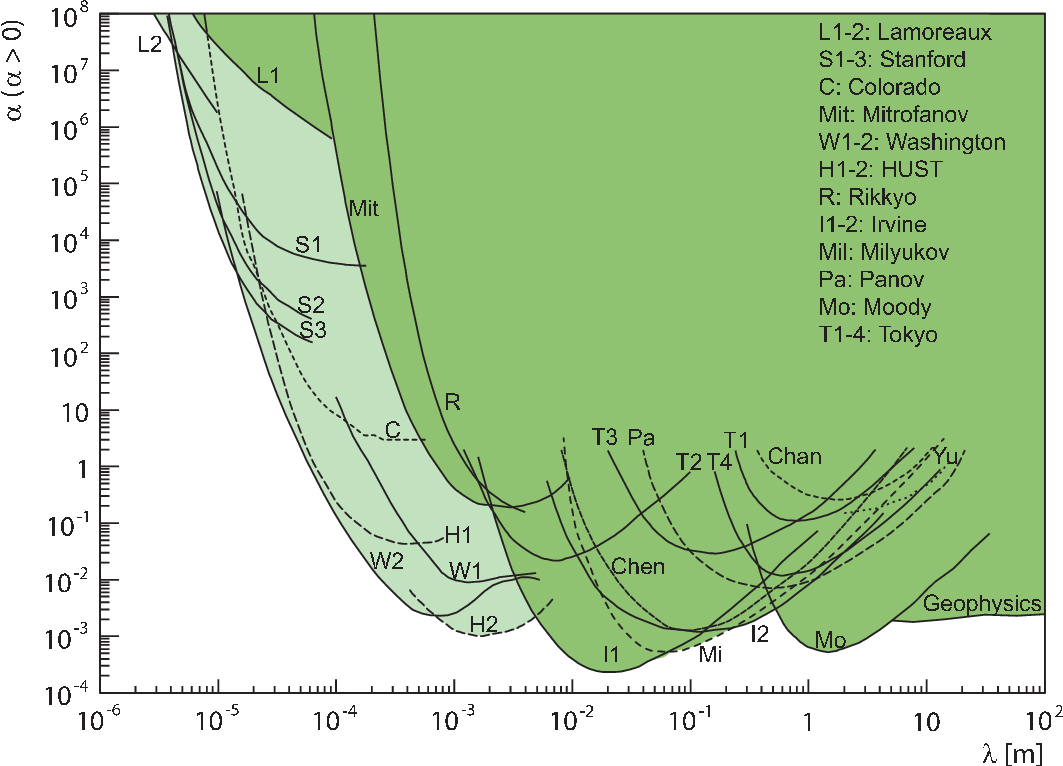}
 \end{center}
 \caption{$\alpha-\lambda$ plot for the laboratory scale experiments. Light shaded area correspond to constraints obtained after the ADD prediction.
Milyukov \cite{Sov-JETP61}; 
Panov \cite{Sov-JETP50}; 
Mitrofanov \cite{Sov-JETP67}; 
Irvine [I1-2]
\cite{PhysRevD.32.3084,PhysRevLett.44.1645}; 
Moody \cite{PhysRevLett.70.1195}; 
Chan \cite{PhysRevLett.49.1745}; 
U. Tokyo [T1-4]
\cite{Nature283,PhysRevD.26.729,PhysRevD.32.342,PhysRevD.36.2321};
Chen \cite{Proc.R.Soc.Lond}; 
U. Washington [W1-2] 
\cite{PhysRevLett.86.1418,PhysRevD.70.042004,PhysRevLett.98.021101}; 
HUST [H1-2] 
\cite{PhysRevLett.98.201101,PhysRevLett.108.081101}; 
Rikkyo \cite{HE_News.32.2014};
Colorado \cite{Nature.01432};
Stanford [S1-3] 
\cite{PhysRevLett.90.151101,PhysRevD.72.122001,PhysRevD.78.022002}; 
Lamoreaux [L1-2]
\cite{PhysRevLett.78.5,PhysRevLett.107.171101}
  .}
 \label{alpha-lambda-shortscale}
\end{figure}

Figure \ref{alpha-lambda-shortscale} summarizes results from nearly all experimental attempts to test the inverse square law at laboratory scales. Most of the experiments that set constraints at $\lambda > 10 \;\mu \rm{m}$ scales were Cavendish-type experiments using torsion balances.
These experiments measure gravitational torques from source masses placed near a test mass attached to torsion balance bars.
The constraint curves are taken from the original publications and were estimated by the original authors.
In Figure \ref{alpha-lambda-shortscale}, all curves correspond to the 95\% confidence level.

\begin{figure}[H]
 \begin{center}
  \includegraphics[width=100mm]{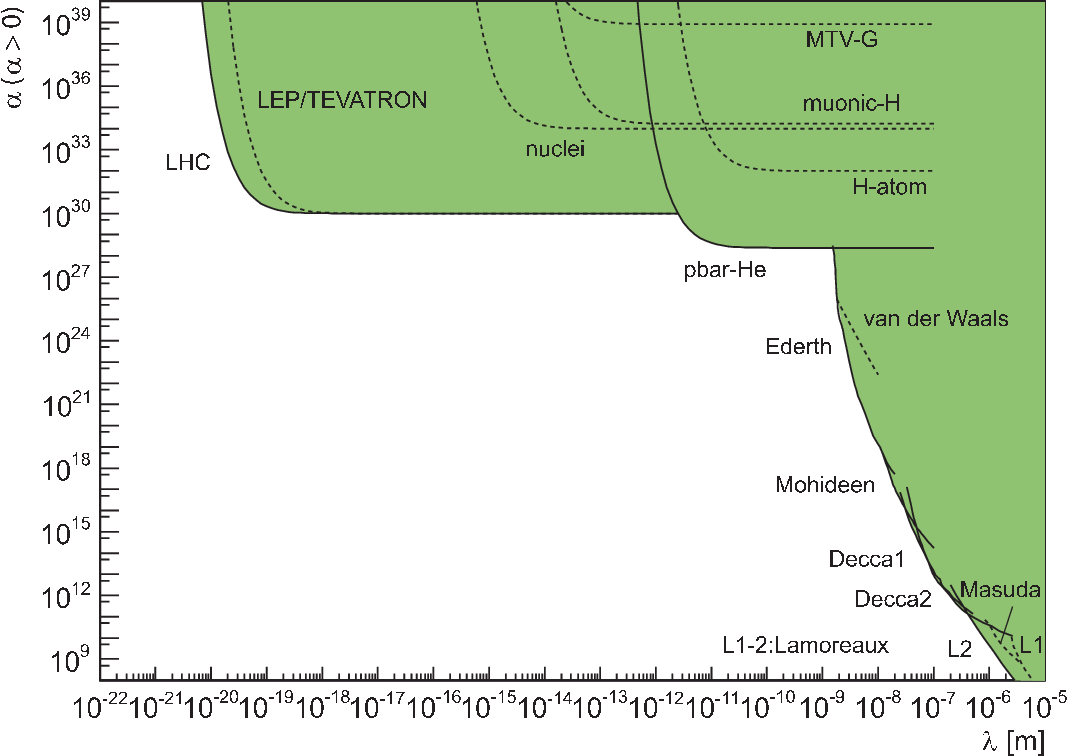}
 \end{center}
 \caption{$\alpha-\lambda$ plot for below $\mu$m scale experiments.
Lamoreaux [L1-2] \cite{PhysRevLett.78.5,PhysRevLett.107.171101}; 
Masuda \cite{PhysRevLett.102.171101}; 
Decca 1-2 \cite{PhysRevLett.94.240401,PhysRevD.75.077101}; 
Mohideen \cite{PhysRevA.62.052109}; 
van der Waals \cite{Israelachvili21111972,Bordag199435}; 
Ederth \cite{PhysRevA.62.062104,PhysRevD.63.115003}; 
H-atom, muonic-H (this analysis);
MTV-G \cite{1742-6596-453-1-012018}; 
pbar-He \cite{INPC2013} and this analysis; 
nuclei  \cite{0954-3899-40-3-035107}; 
LEP \cite{arXiv:hep-ex0410004},
TEVATRON \cite{PhysRevLett.101.181602},
LHC \cite{PhysRevLett.110.011802,Aad2011294} and this analysis.
 }
 \label{alpha-lambda-microscale}
\end{figure}

For experiments performed below $\mu$m scales, results are shown in Figure \ref{alpha-lambda-microscale}. Experiments in $\lambda = 10 \sim 100 \; \mu \rm{m}$ scales used cantilevers as the force sensor instead of torsion balances. Such microscale experiments were performed as gravitational resonance searches at the moving frequency of the source.
Experimental sensitivities are far above Newtonian gravity; therefore, the resulting constraints on $\alpha$ are very big numbers at $\alpha\gg 1$.

At distance scales below $10 \; \mu \rm{m}$, experiments measuring the Casimir force were used as the input data to constrain the strong gravity.
At this scale, electromagnetic shielding is difficult to place between the source and test masses.
Therefore, the dominant forces on the system are the remaining direct electromagnetic forces.
Some curves were obtained by independent authors using published experimental data.
Especially for the Casimir force measurements, theoretical estimates are themselves complex.

In Figure \ref{alpha-lambda-microscale}, results from atomic, nuclear, and particle physics data are also plotted.
Just as for the Casimir force region, the dominant forces on these systems are known standard model interactions, such as electromagnetic interactions, as well as strong and weak interactions.
Therefore, these constraints are estimated to be the maximum allowed strength of strong gravity within the experimental precision of the standard model interactions.
Most of these constraints on $\alpha-\lambda$ parameter space (hydrogen atom, antiprotonic (pbar) helium atom, muonic hydrogen, LEP/TEVATRON/LHC) were obtained in the present study, as described in Section \ref{subsec-atomic}.
Note that the results from high-energy collider experiments assume graviton-producing phenomena; in contrast, other nuclear and atomic data were analyzed by assuming classical gravitational phenomena.
Therefore, the theoretical reliabilities used in the evaluations are very different.

\subsection{Constraints on the power-law parametrization}

\begin{figure}[t]
 \begin{center}
  \includegraphics[width=100mm]{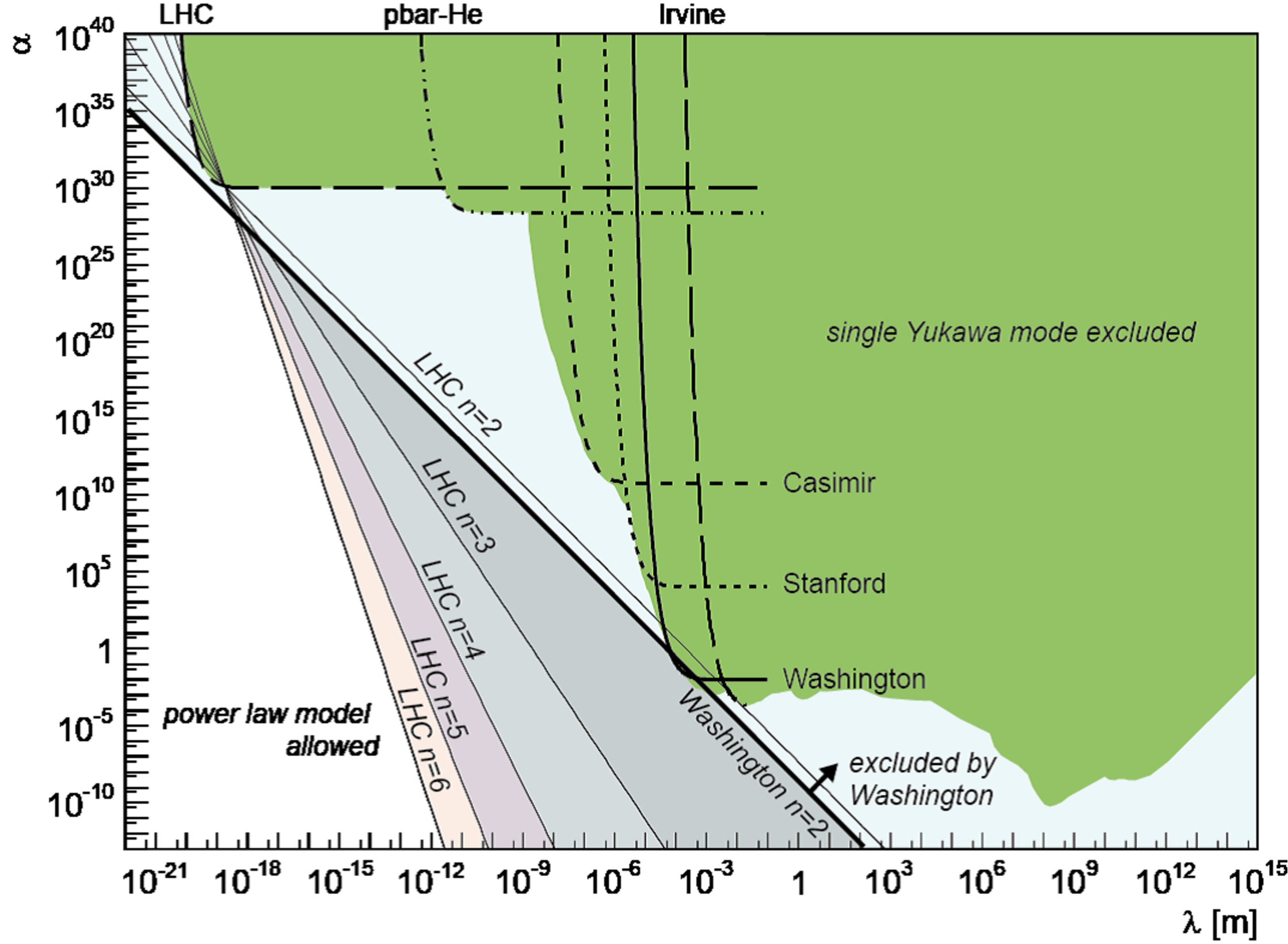}
 \end{center}
 \caption{$\alpha-\lambda$ plot showing typical curves representing important kink points, going to be used in $n-\lambda$ plot after interpretation via corresponding $\delta$. 
Representing curves are set around region of Irvine, Washington, Stanford, Casimir. Curves for pbar-He and LHC curves are set exactly using the present analysis.
The curves are drawn using single exponential formula of equation (\ref{alpha-Yukawa-abs}). 
 Straight slopes showing modified $\alpha^{power}_{n}(\lambda)$ (equation (\ref{alpha_n})), are drawn for the LHC ($n=2,3,4,5,6$) and for the Washington ($n=2$).}
 \label{alpha-lambda-explain}
\end{figure}

\begin{figure}[H]
 \begin{center}
  \includegraphics[width=100mm]{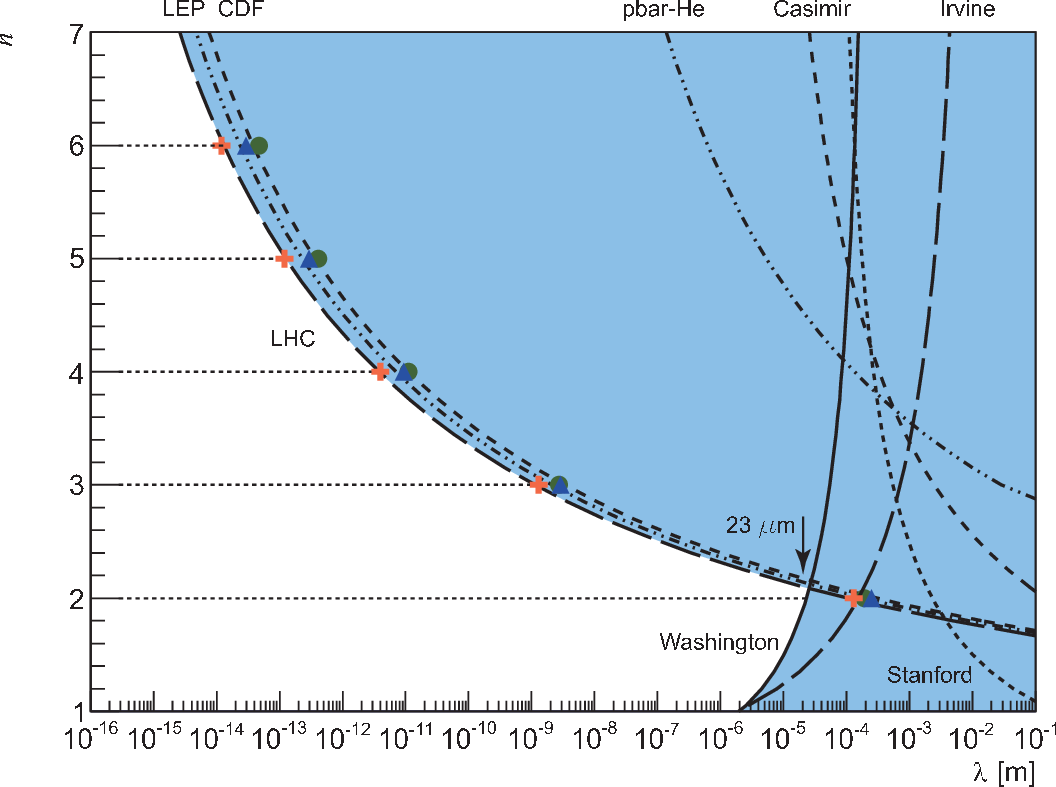}
 \end{center}
 \caption{$n-\lambda$ plot for representing Irvine, Washington, Stanford, Casimir, pbar-He and LEP/CDF/LHC results, which are corresponding to that in Figure \ref{alpha-lambda-explain}. The $\lambda$ values of the collider results are calculated using the ADD model.
Marked points are obtained for LEP \cite{arXiv:hep-ex0410004}, TEVATRON \cite{PhysRevLett.101.181602}, LHC \cite{PhysRevLett.110.011802,Aad2011294}. Curves are drawn using equation (\ref{n-power}). Shaded area (large $\lambda$) are excluded regions. For $n=2$, the Washington data sets the best limit $\lambda < 23 \;\mu\rm{m}$.}
 \label{n-lambda}
\end{figure}

Constraints on the power-law parameter space have not been reported for most experiments, except collider experiments, which did not report results for the Yukawa parametrization.
To set experimental constraints on the $n-\lambda$ parameter space in the power-law parametrization without using raw experimental data, such as $\delta$, several typical constraint curves on the $\alpha-\lambda$ space were chosen, as shown in Figure \ref{alpha-lambda-explain}.
These representative curves were selected to reproduce the important kink points (for Irvine, Washington, Stanford, Casimir, pbar-He, and LHC), using the point-mass formula in (\ref{alpha-Yukawa}).
Then, the corresponding values of $\delta$ for these representative curves can be obtained;
this enables us to interpret the constraints in the Yukawa form in terms of the power-law form.
The results are shown in Figure \ref{n-lambda}.
Results from collider experiments were obtained assuming the ADD model; this allows us  to interpret the results obtained as ``$n$ vs $M_D$" in the form of $\alpha-\lambda$.
The upper limits on $\lambda$ were obtained from the Washington data for $n=2$ and from the LHC data for $n\geq 3$.


\begin{figure}[H]
 \begin{center}
  \includegraphics[width=100mm]{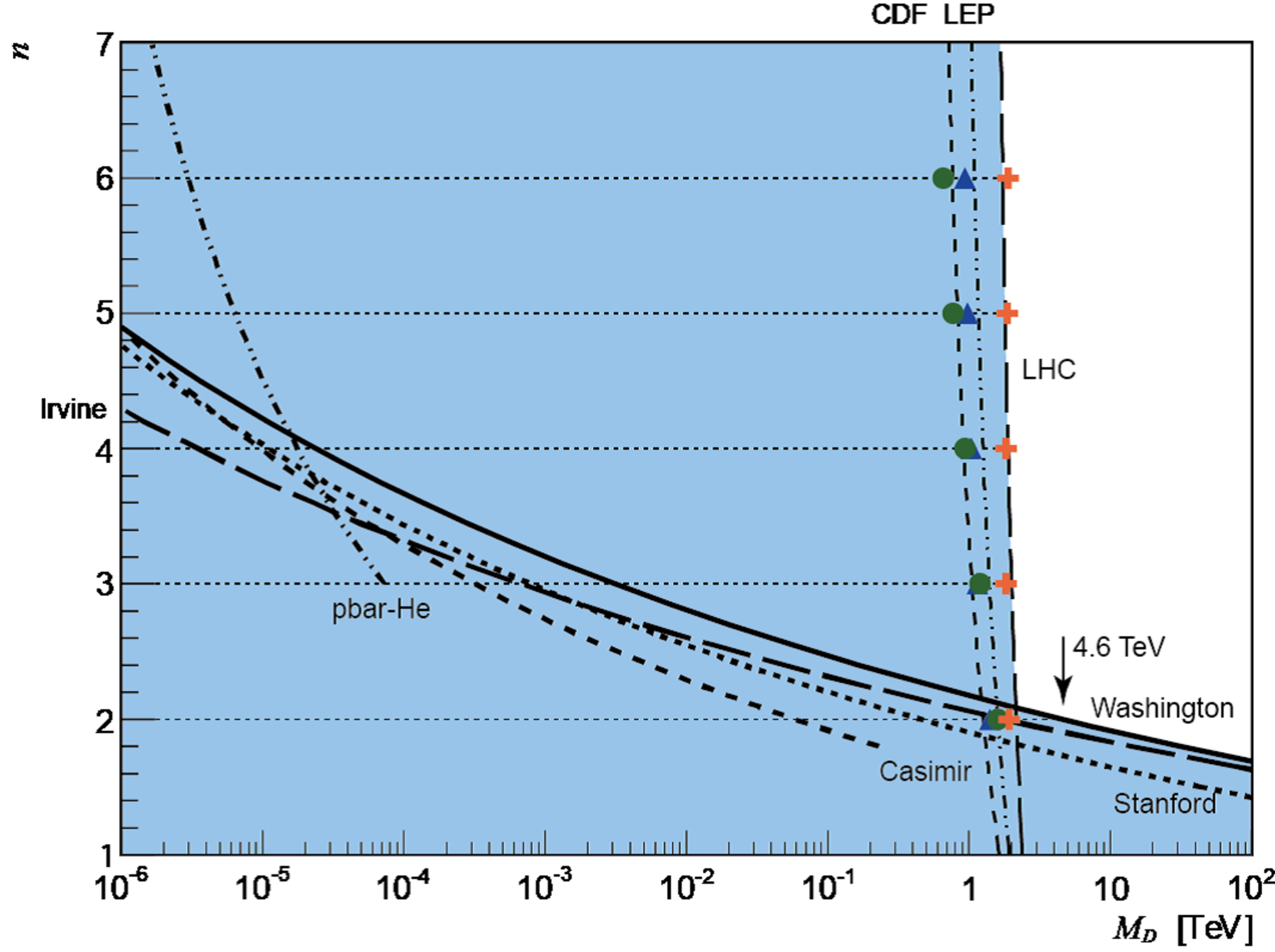}
 \end{center}
 \caption{$n-M_D$ plot for representing Irvine, Washington, Stanford, Casimir, pbar-He and LEP/CDF/LHC results, which are corresponding to that in Figure \ref{alpha-lambda-explain}. The marked $M_D$ values of the collider results are obtained in one jet production cross section at LEP \cite{arXiv:hep-ex0410004}, TEVATRON \cite{PhysRevLett.101.181602}, LHC \cite{PhysRevLett.110.011802,Aad2011294}. Curves are drawn using equation (\ref{n-power}) combined with equation (\ref{ADD-lambda-MD}). Shaded area (small $M_D$) are excluded regions. For $n=2$, the Washington data sets the best limit $M_D > 4.6 \;\rm{TeV}$.}
 \label{n-MD}
\end{figure}

\section{Discussion on the ADD model}
\label{add-discussion}
To compare results from the short-range experiments and those from collider experiments, it is useful to express the results in $n-M_D$ parameter space.
The relation between $\lambda$ and $M_D$ was obtained assuming the ADD model
\begin{equation}
\lambda=\frac{(M_{pl}/\sqrt{8\pi})^{2/n}}{M_D^{1+2/n}} \frac{\hbar}{c},
\label{ADD-lambda-MD}
\end{equation}
where the Planck mass $M_{pl}/\sqrt{8\pi}=\sqrt{\hbar c/\kappa^2}=2.44 \times 10^{15} \;\rm{TeV/c}^2$ was estimated using Einstein's gravitational constant 
$\kappa^2 = 8\pi G_N$ instead of Newtonian gravitational constant $G_N$.
Figure \ref{n-MD} shows the resulting $n-M_D$ plot that corresponds to the $n-\lambda$ plot in Figure \ref{n-lambda}.
Moreover, the experimental lower limits on $M_D$ were determined from the Washington data for $n=2$ and from the LHC data for $n\geq 3$.
The results from the Washington data at $n=2$ are interpreted as $M_D > 4.6 \;\rm{TeV}$ and $\lambda < 23 \;\mu \rm{m}$.
This constraint is stronger than that in the original publication by the University of Washington group \cite{PhysRevLett.98.021101} wherein corresponding values were estimated from an $\alpha-\lambda$ plot by searching over $\lambda$ at $\alpha=8/3$ \cite{Ann.Rev.53} and assuming Yukawa potential approximation; the latter is valid if the experiment is performed around $\lambda$.
As discussed in Section \ref{section-intro}, the ADD model is  primarily expressed in a power-law functional form.
However, the conventional Yukawa form can only be applied near $\lambda \sim r$ by assuming that lowest order diagram exchanges the lightest graviton.
Indeed, interpretation of $n$ in the ADD model in terms of the Yukawa parameter $\alpha$ is possible, as has been calculated by some authors
 \cite{Kehagias200039,Ann.Rev.53}.
For a very small experimental distance $r\ll\lambda$, $G(r)$ must become very large due to the power-law form.
However, such magnification at very short distances cannot be represented by constant $\alpha$ parameter in the Yukawa parametrization. 
In fact, $\alpha$ should be modified if the Yukawa parametrization is applied to analyze experimental data obtained at $r\ll \lambda$.
Setting $a_{Yukawa}(r) = a_{power}(r)$ indicates that physical phenomena obey a power-law-type dependence, but for experimental data analyzed using the Yukawa form,
a modified $\alpha$ can be expressed by
\begin{equation}
\alpha'(\lambda)=\frac{(1+n)\left( \frac{\lambda}{r}\right)^n}{(1+\frac{r}{\lambda})e^{-r/\lambda}} \rightarrow (1+n)\left( \frac{\lambda}{r}\right)^n \;\; (r/\lambda \rightarrow 0),
\end{equation}
for point-mass calculations.
This means that $\alpha$ seems to be amplified at a small experimental scale $r \ll \lambda$, if the actual gravitational potential obeys the power-law form.
To obtain the original strength $\alpha$ at $r\sim \lambda$, the amplification factor must be removed.
The amplification factor is
\begin{equation}
A(\lambda)=\frac{\alpha'(\lambda)}{\alpha'(\lambda=r)}=\frac{\left( \frac{\lambda}{r}\right)^n}{(1+\frac{r}{\lambda})}2e^{r/\lambda-1},
\end{equation}
and a corrected $\alpha$ for the simple exponential function in (\ref{alpha-Yukawa-abs}) can be obtained
\begin{equation}
\alpha^{power}_n(\lambda)=\frac{\alpha(\lambda)}{A(\lambda)}=\frac{e}{2} \;\delta \left( \frac{r}{\lambda}\right)^n,
\label{alpha_n}
\end{equation} 
which is not constant but a function of $n$ and $r$.
The resulting corrected $\alpha^{power}_n$ is plotted in Figure \ref{alpha-lambda-explain} for the LHC data in $n=2 - 6$, and for the Washington data at $n=2$.
It is now clear that the experimental sensitivities should be compared with the inclined lines $\alpha_{n}^{power}$ for each $n$.
For $n=2$, the slope is $\lambda^{-2}$.
To compare the sensitivities in the ADD model on the $\alpha-\lambda$ plot, the constraint curves need to be rotated until the corrected lines $\lambda^{-n}$ become horizontal.
Then, the Washington kink is the deepest valley in the vertical direction for this $n=2$ line.
Figure \ref{alpha-lambda-explain} clearly shows that the tightest constraints are set by the LHC data for $n\geq 3$ and that the $\mu \rm{m}$ scale is the best position to  constraint for $n=2$.

\section{List of Experiments}
In this section, we briefly introduce the experiments that provided the data used in Figures \ref{alpha-lambda-shortscale} and \ref{alpha-lambda-microscale}. All the experiments involved test and source masses; the forces between them were measured.
In a laboratory experiment using a torsion balance, the effects of other forces can be reduced by placing an electric shield between the test and source masses, so that the measured force can be regarded as being dominated by gravity. Then, the suppression of Newtonian gravity helps test the inverse square law. In a microscopic measurement, the gravitational force is hidden in the dominant background from Coulomb forces, including the Casimir force. Therefore, only an upper limit on the non-Newtonian strong gravity can be determined. Atomic and sub-atomic systems can be regarded as test systems similar to microscopic measurements wherein standard model backgrounds cannot be avoided.

\subsection{Torsion balance experiments}
\begin{figure}[H]
 \begin{center}
  \includegraphics[width=90mm]{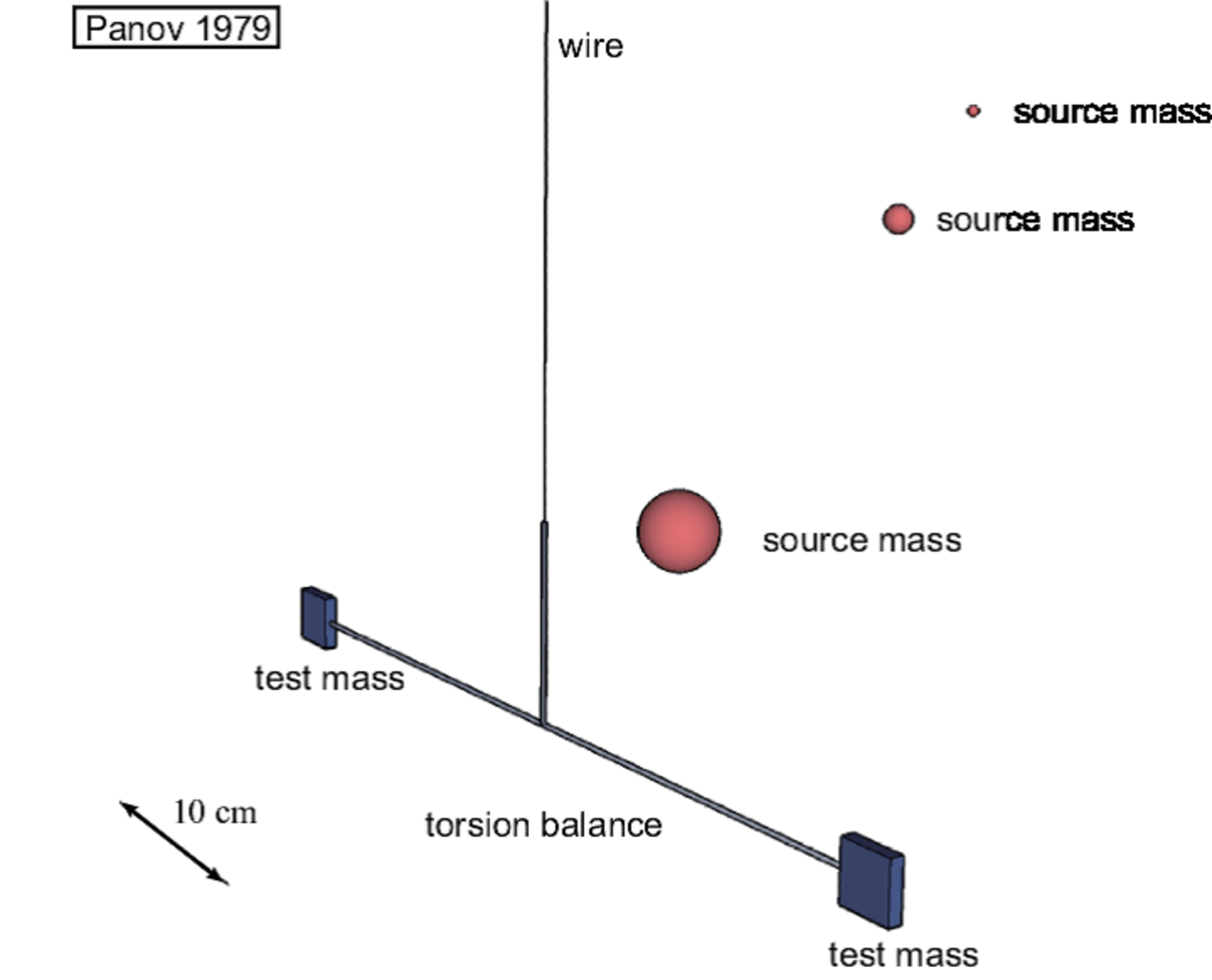}
 \end{center}
 \caption{Experiment of Panov at Moscow State University \cite{Sov-JETP50}. Test mass (quartz) $m=10$ g. Source masses : ($r_0=0.4$ m, $M_0=0.2$ kg), ($r_1=3.0$ m, $M_1=56$ kg), ($r_2=9.8$ m, $M_2=595$ kg) (source mass center to balance bar center distance). 
Torsion balance bar : (total length) $2L = 40$ cm.  
 Sensor : capacitive displacement sensor. Wire : tungsten 30 $\mu$m diameter, 31 cm long.  Torsion balance is in a glass vacuum chamber.}
 \label{Panov}
\end{figure}

\begin{figure}[H]
 \begin{center}
  \includegraphics[width=70mm]{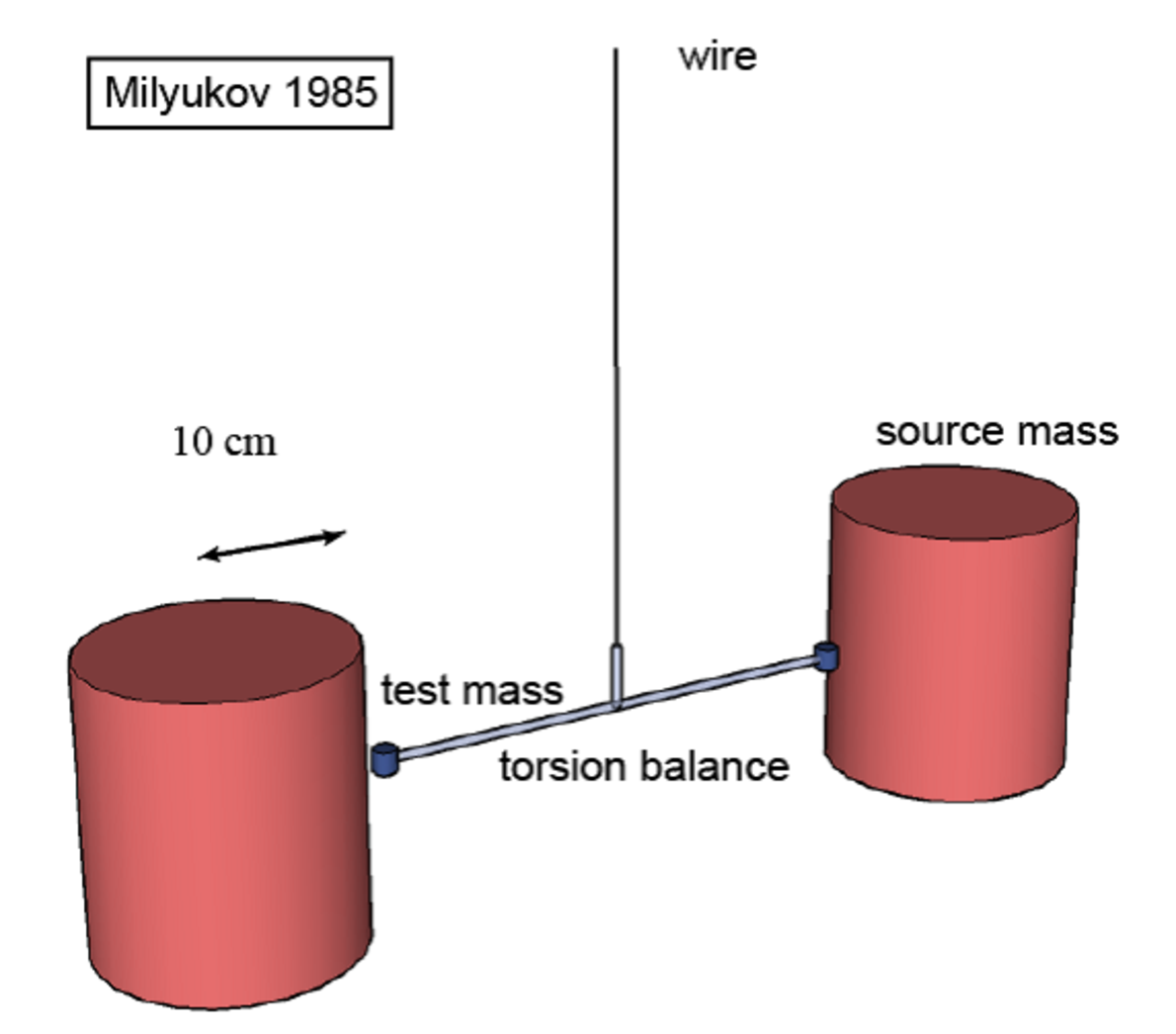}
 \end{center}
 \caption{Experiment of Milyukov at P.K. Shternberg State Astronomical Institute \cite{Sov-JETP61}. Test mass (copper cylinder) $m=29.9$ g. Source masses (18 cm diameter, 20 cm height nonmagnetic steel cylinder) : $M=40$ kg at $r=11.25, 13.25, 16.25, 21.25$ cm (center to center distances). Torsion balance bar (total length) $2L$ = 35.5 cm. Sensor : optical lever displacement sensor. Wire : tungsten $1 \mu$m [sic] diameter, 1 m long. Frequencies of the torsional vibration were observed. Torsion balance is in a copper vacuum chamber.}
 \label{Milyukov}
\end{figure}

\begin{figure}[H]
 \begin{center}
  \includegraphics[width=62mm]{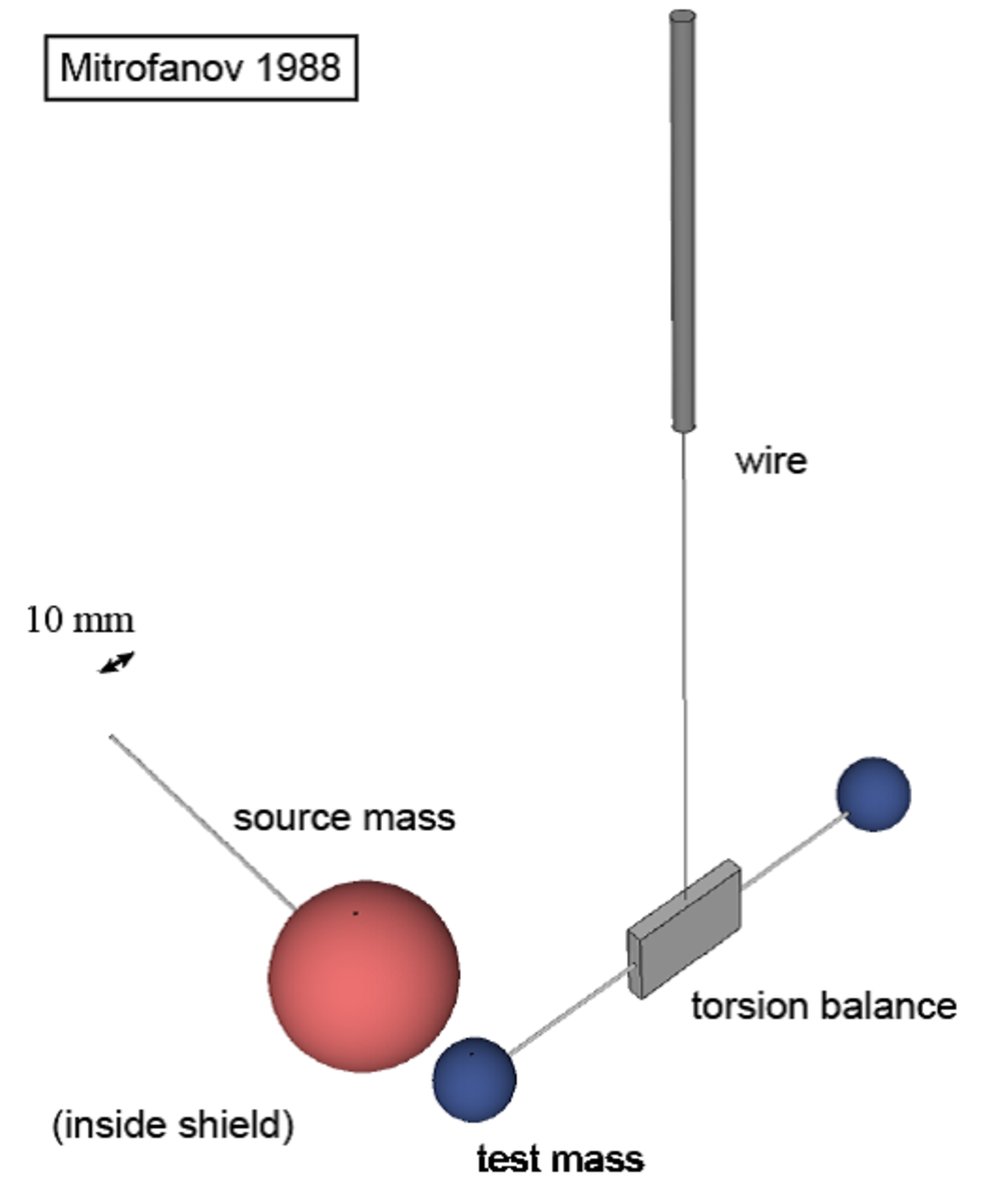}
 \end{center}
 \caption{Experiment of Mitrofanov at M.V. Lomonosov State University \cite{Sov-JETP67}. Test mass (platinum sphere) $m=59$ mg. Source mass (tungsten sphere) : $M=706$ mg at $r_1 = 3.8$ mm and $r_2 = 6.5$ mm (center to center distance of the spheres). Sensor : capacitive parametric displacement sensor. Wire : aluminized quartz $5 \mu$m diameter, 14 mm long. Torsion balance is in a 0.25 mm thick vacuum chamber.}
 \label{Mitrofanov}
\end{figure}

\begin{figure}[H]
 \begin{center}
  \includegraphics[width=80mm]{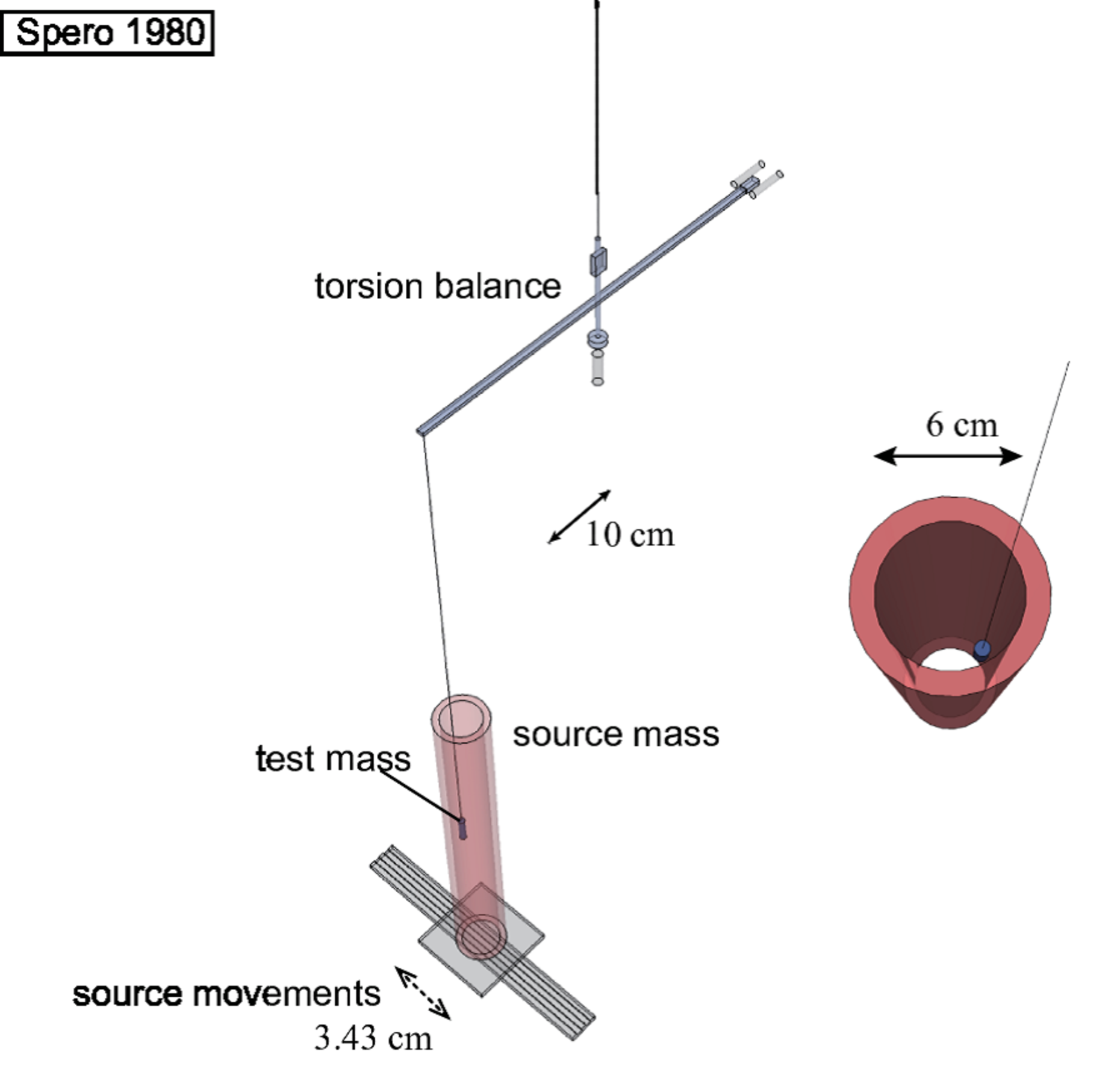}
 \end{center}
 \caption{Experiment of Spero at University of California, Irvine \cite{PhysRevLett.44.1645}.
 Test mass (4.4 cm height copper cylinder): $m=20$ g. Source mass (inner diameter 6 cm, outer diameter 8 cm, 60 cm height stainless steel pipe): $r=$ 1.8, 5.2 cm, $M =$ 10.44 kg (test mass center to pipe wall radial center). Torsion balance bar (total length) $2L$ = 60 cm. Sensor : optical lever displacement sensor. Wire : 75 $\mu$m diamater, 32 cm long tungsten. 
Torsion balance position is stabilized using feedback on electrostatic force plates.
Torsion balance is in concentric copper vacuum chamber.
Source mass moves reciprocating along the horizontal rail.
}
 \label{Spero}
\end{figure}

\begin{figure}[H]
 \begin{center}
  \includegraphics[width=120mm]{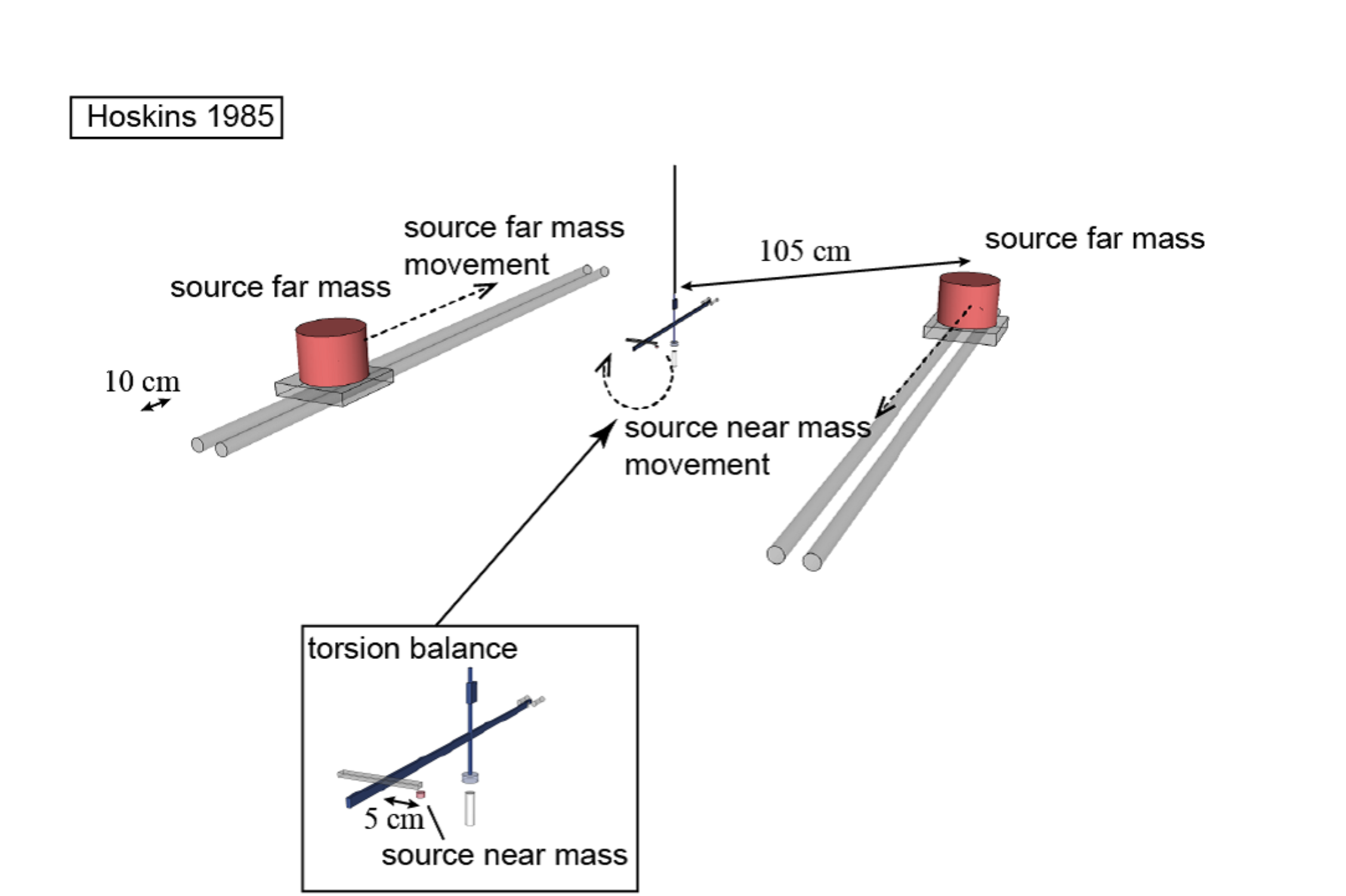}
 \end{center}
 \caption{Experiments of Hoskins at University of California, Irvine \cite{PhysRevD.32.3084}. 
 Test mass = torsion balance bar : $m$ = 523 g, 60 cm long copper bar. Source masses (copper cylinder): $r = $5 cm, $M = $43 g  (near mass center to bar center), $r =$105 cm, $M =$ 7.3 kg (far source mass center to balance bar center). Wire : tungsten 90 $\mu$m diameter, 20 cm long. Sensor : optical lever displacement sensor. Torsion balance position is stabilized using feedback on electrostatic force plates. 
Torsion balance is in vacuum chamber surrounded by magnetic and thermal shield.
Source near mass moves reciprocating over opposite side of torsion balance.
Source far mass moves reciprocating along the horizontal rail.
 }
 \label{Hoskins}
\end{figure}

\begin{figure}[H]
 \begin{center}
  \includegraphics[width=70mm]{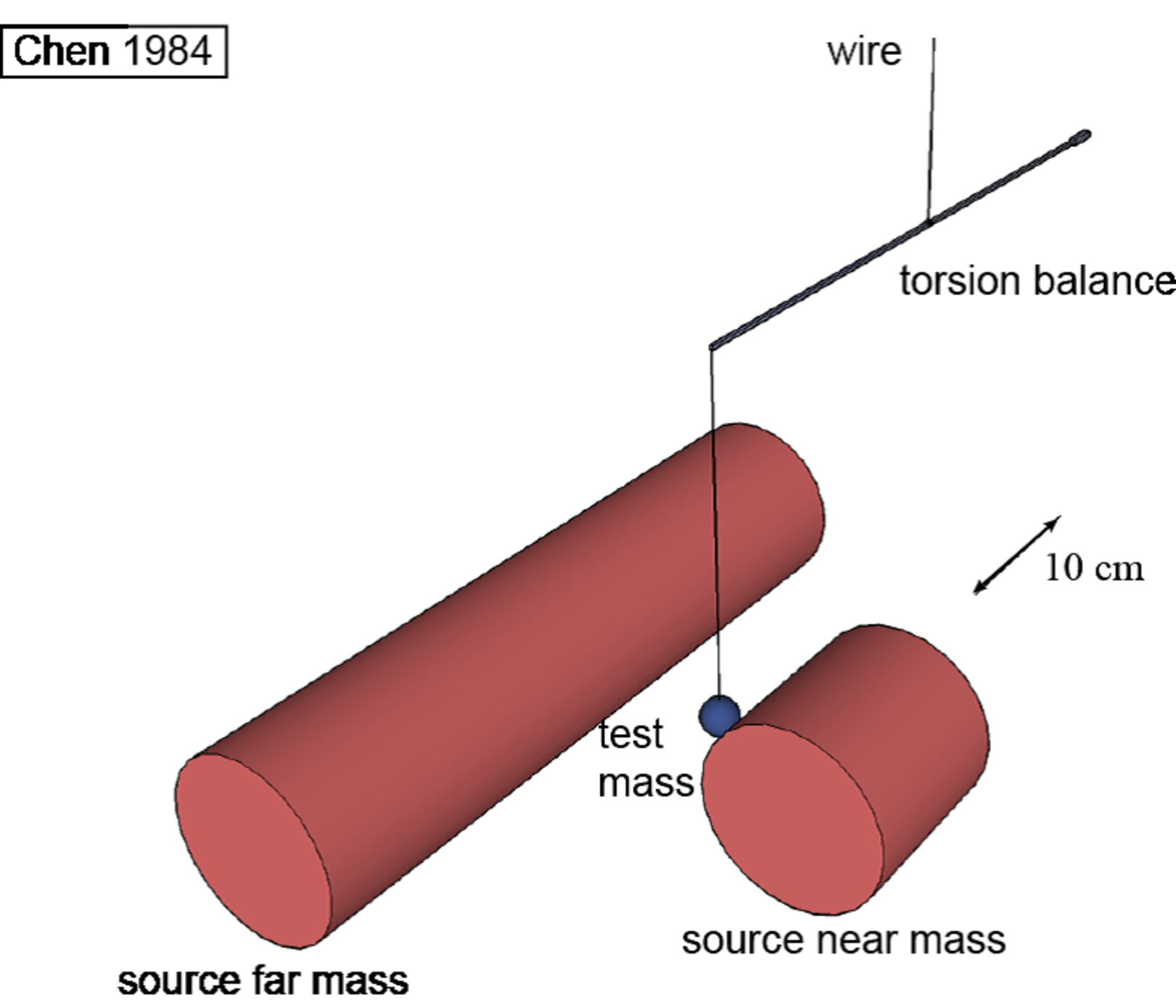}
 \end{center}
 \caption{Experiment of Chen at Cavendish Laboratory \cite{Proc.R.Soc.Lond}. 
 Test mass (phosphor-bronze sphere): 41 g. Source masses (non-magnetic stainless steel cylinder, 10 cm diameter):  $r_1 = 5$ cm (center to center distances), $M =$ 6 kg (10 cm height) and $r_2 =$ 9 cm, $M =$ 25 kg (40 cm height). Torsion balance bar : 60 cm. Wire : tungsten 75 $\mu$m diameter, 80 cm long. Sensor : optical lever displacement sensor.
Torsion balance is in glass vacuum chamber.
 }
 \label{Chen}
\end{figure}

\begin{figure}[H]
 \begin{center}
  \includegraphics[width=130mm]{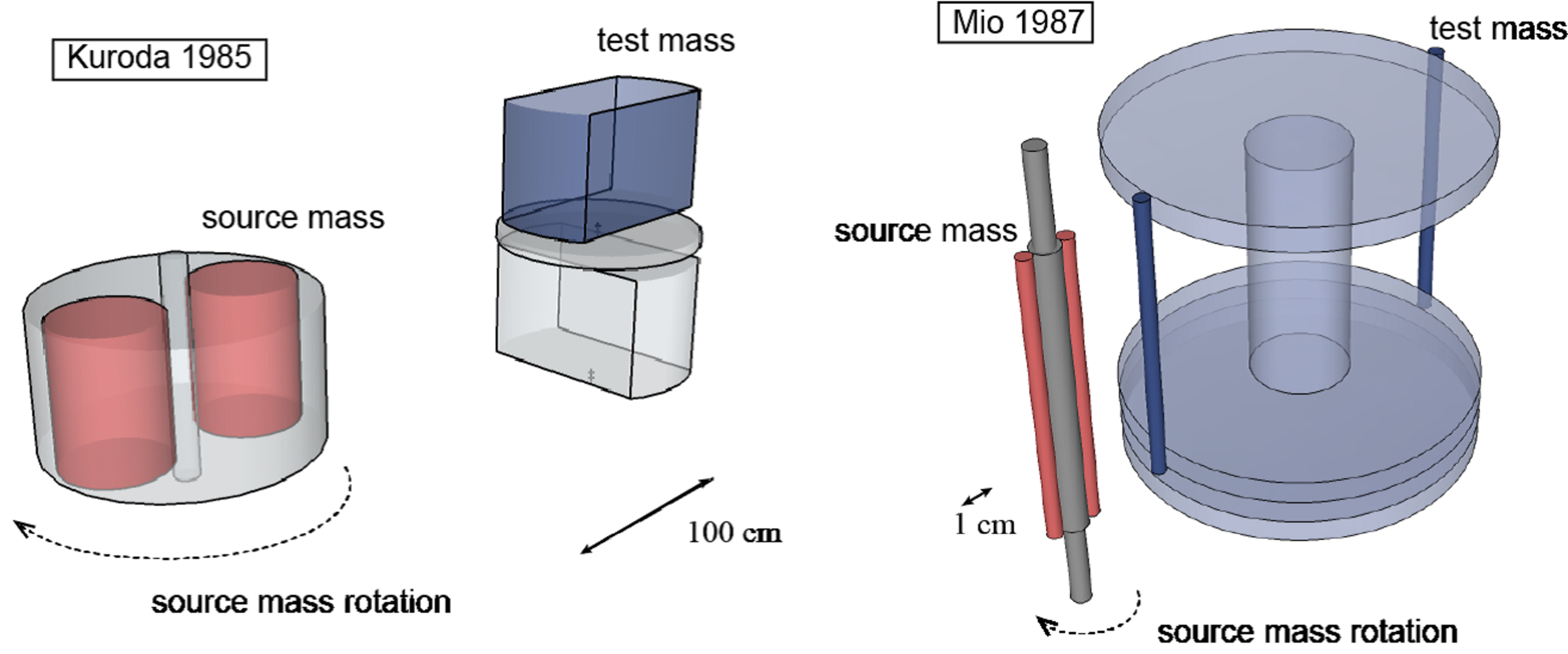}
 \end{center}
 \caption{Experiments of the University of Tokyo group. 
 [Left]: Experiment by Kuroda \cite{PhysRevD.32.342}.
Test mass (gravity wave antenna, side cut aluminum cylinder) 0.85 kg for 10 - 15 cm (distance between source and target rotation centers), and 15 kg for 15 - 30 cm. Source masses (gravity wave radiator, 35 mm diameter, 45 mm height lead cylinders): 0.49 kg. 
[Right]: Experiment by Mio \cite{PhysRevD.36.2321}.
Test mass (gravity wave antenna) 5 mm diameter, 100 mm long tungsten cylinders at 62.5 mm (radial distance from rotation center). 
Source masses (gravity wave antenna): 5 mm diameter, 100 mm long tungsten cylinders at 7.1 mm (radial distance from rotation center), 
at 78 - 87 mm (distance between source and target rotation centers).
Source mass rotates around the vertical axis.
 }
 \label{Kuroda-Mio}
\end{figure}

Most experimental tests of the inverse square law have been performed using Cavendish's torsion balances.
These measurements were triggered by Long's experimental claim for a violation of the inverse square law \cite{Nature260}; this claim was published after his reanalysis of existing data \cite{PhysRevD.9.850}, which was indicating a violation of the inverse square law.
Subsequently, Soviet groups performed a few important precision measurements at cm scale using classic torsion balance bars:
(Figure \ref{Panov}) with $\alpha<7\times 10^{-3}$ at $\lambda=$ 0.4 m \cite{Sov-JETP50}, 
(Figure \ref{Milyukov}) with $\alpha<5\times 10^{-4}$ at $\lambda=$ 6 cm \cite{Sov-JETP61}, 
and (Figure \ref{Mitrofanov}) with $\alpha<2\times 10^{-1}$ at $\lambda=$ 2 mm \cite{Sov-JETP67}. 
In the 1980s, the University of Irvine group performed a couple of very accurate measurements using static torsion balance bars, which to date still set the tightest constraints at cm scale:
(Figure \ref{Hoskins}) with $\alpha<1\times 10^{-3}$ at $\lambda=$ 10 cm \cite{PhysRevD.32.3084}, 
(Figure \ref{Spero}) with $\alpha<2\times 10^{-4}$ at $\lambda=$ 2 cm \cite{PhysRevLett.44.1645}. 
The Cavendish laboratory group used a torsion balance bar
(Figure \ref{Chen}) with $\alpha<1\times 10^{-3}$ at $\lambda=$ 10 cm \cite{Proc.R.Soc.Lond}. 
The University of Maryland group performed experiments without using torsion balances but utilizing superconducting gravity gradiometers
with $\alpha<1\times 10^{-1}$ at $\lambda=$ 80 cm \cite{PhysRevLett.49.1745}, 
and with $\alpha<5\times 10^{-4}$ at $\lambda=$ 1 m \cite{PhysRevLett.70.1195}, 
which still gives the best limit at the 1 m scale.
Similarly, Goodkind also performed a superconducting gravimeter experiment 
with $\alpha<1\times 10^{-2}$ at $\lambda=$ 1 m \cite{PhysRevD.47.1290}. 
The University of Tokyo group performed a series of experiments utilizing gravity wave antenna:
with $\alpha<3\times 10^{-1}$ at $\lambda=$ 2 m \cite{Nature283}, 
with $\alpha<1\times 10^{-2}$ at $\lambda=$ 50 cm \cite{PhysRevD.26.729}, 
(Figure \ref{Kuroda-Mio}) with $\alpha<3\times 10^{-2}$ at $\lambda=$ 10 cm \cite{PhysRevD.32.342}, 
and (Figure \ref{Kuroda-Mio}) with $\alpha<2\times 10^{-2}$ at $\lambda=$ 7 mm \cite{PhysRevD.36.2321}. 
A similar gravity wave antenna experiment was performed at CERN
\cite{astone1991}.
All these results are consistent with the inverse square law.
The most significant result is that from the Irvine group.

\begin{figure}[t]
 \begin{center}
  \includegraphics[width=70mm]{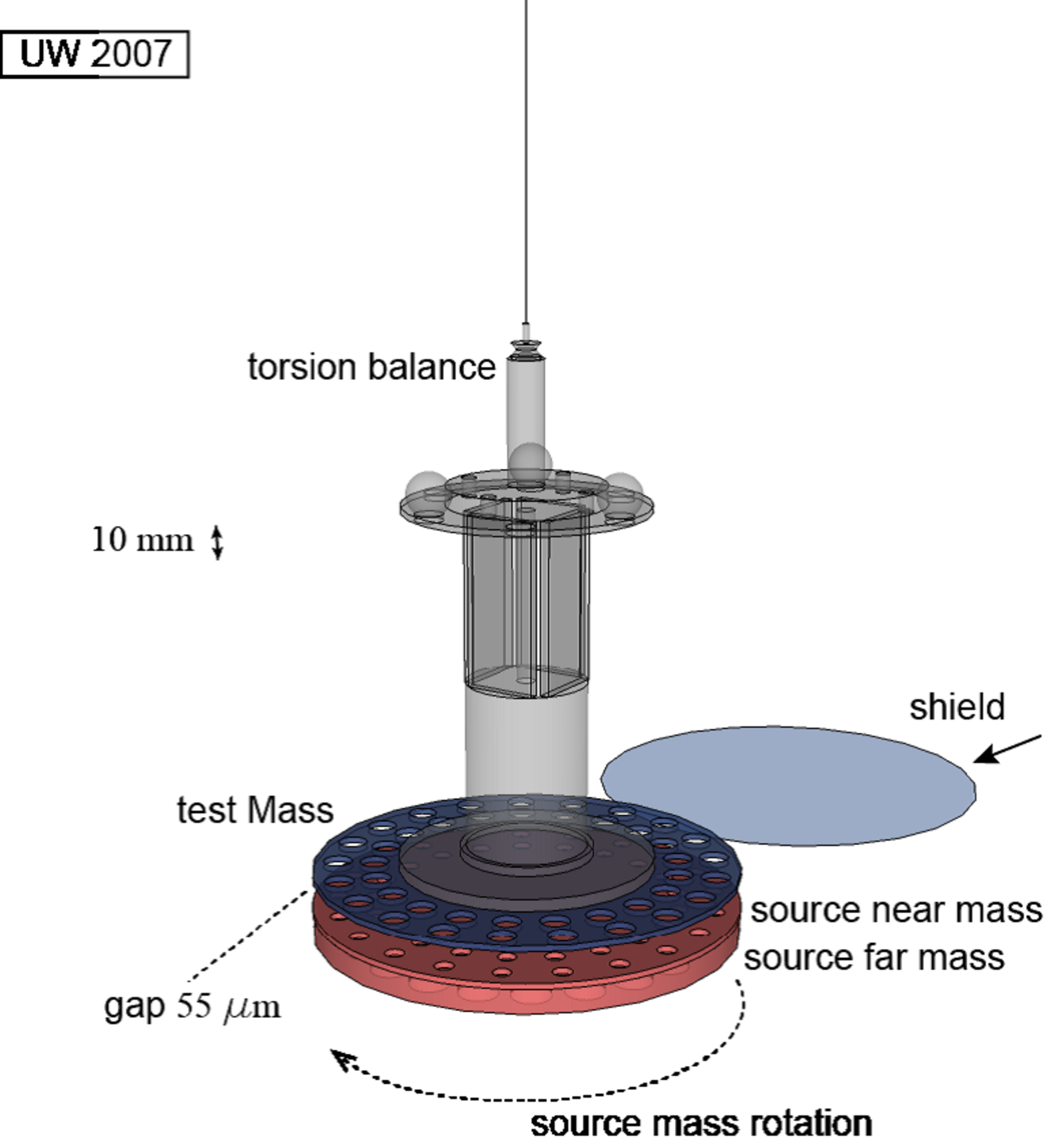}
 \end{center}
 \caption{Experiment of University of Washington group 
 \cite{PhysRevLett.98.021101}. 
Test mass (42 of 1 mm thick, 4.8 mm diameter holes in molybdenum disk), Source near masses (42 of 1 mm thick, 3.2 mm diameter holes in molybdenum disk) at $d_{near} =$ 55 $\mu$m - 9.53 mm (surface to surface) gap. Source far masses (42 of 3 mm thick, 6.4 mm diameter holes in tantalum disk) : attached below near mass. Sensor : Optical lever. Wire : tungsten 20 or 17 $\mu$m diameter, 80 cm long. 
The torsion pendulum is not stabilized, but the dynamic motion is monitored.
Electric shield (10 $\mu$m gold coated beryllium copper membrane) is set between test and source near mass.
Source mass rotates around the vertical axis.
}
 \label{UW}
\end{figure}

The second impact on this field was the proposal of the large extra-dimension model
\cite{Arkani.Hamed1998263}.
The striking prediction of a possible violation of the inverse square law at around 0.1 mm triggered many recent sophisticated experiments.
Before then, the tightest constraint was given by the result of Mitrofanov \cite{Sov-JETP67}.
The University of Washington group (E\"ot-wash) obtained a very strong constraint $\alpha<9\times 10^{-3}$ at $\lambda=$ 1 mm using a torsion pendulum with missing mass holes on a test disk
\cite{PhysRevLett.86.1418,PhysRevD.70.042004}.
They also obtained an upgraded result
(Figure \ref{UW}) with $\alpha<2\times 10^{-3}$ at $\lambda=$ 0.5 mm \cite{PhysRevLett.98.021101}.
Many results from the E\"ot-wash group were introduced in \cite{Adelberger:2009zz}, not only for the inverse square law but also for other gravity-related measurements.


\begin{figure}[H]
 \begin{center}
  \includegraphics[width=75mm]{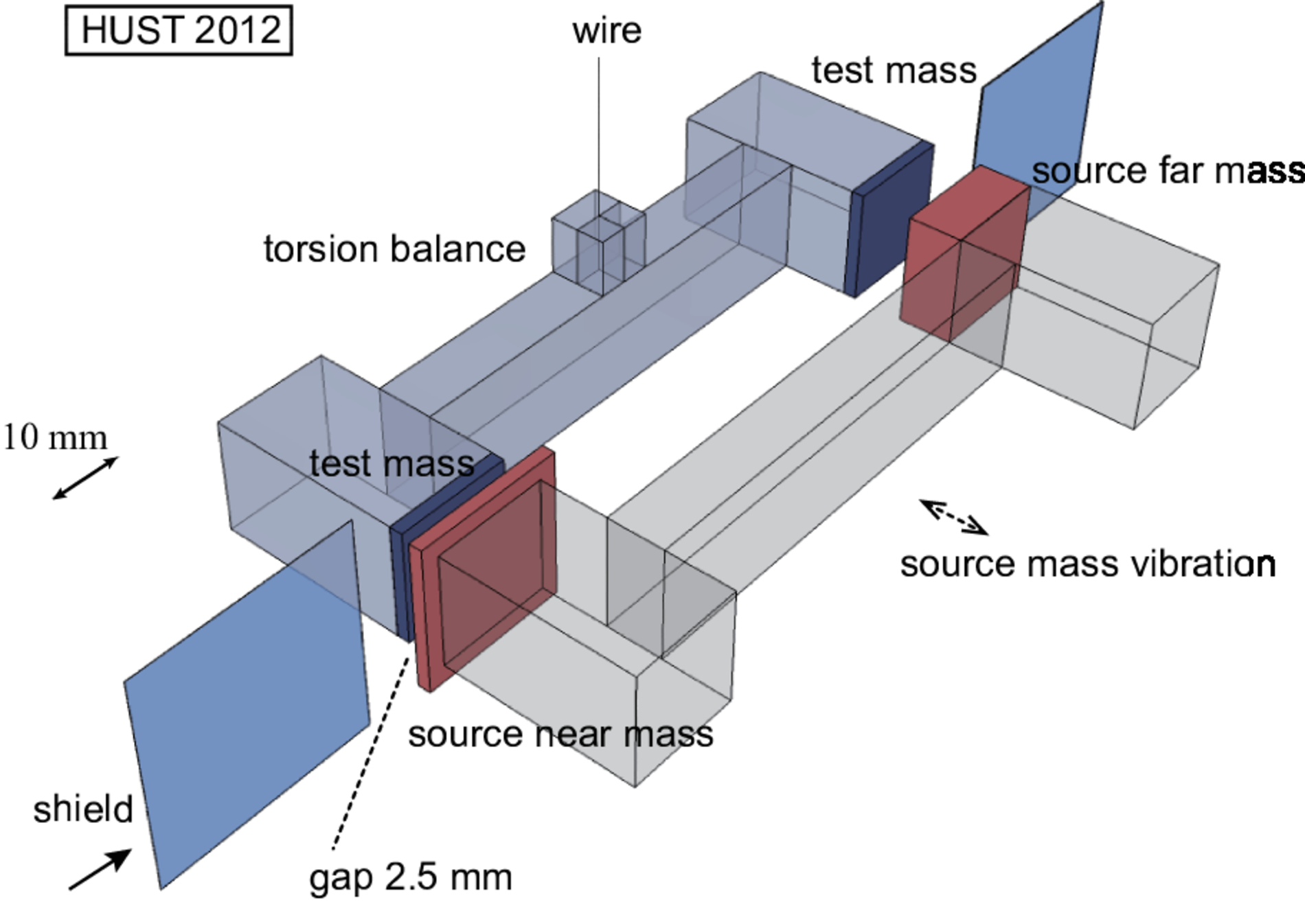}
 \end{center}
 \caption{Experiment of the HUST group \cite{PhysRevLett.108.081101}, \cite{PhysRevLett.98.201101}. 
Test mass : 1.8 mm thick, 16 mm $\times$ 16 mm wide tungsten plate. Source near mass : 1.8 mm thick, 20.8 mm $\times$ 20.8 mm wide tungsten plate at $d_{near} =$ 0.7 mm (surface to surface) gap. 
Source far mass : 7.6 mm thick, 16 mm $\times$ 16 mm wide tungsten plate at $d_{far} =$ 4 mm (surface to surface) gap.
Torsion balance bar (end to end length) $2L$ = 100 mm.
Sensor : autocollimator. Wire : tungsten $25 \mu$m diameter, 60 cm long. 
Torsion balance position is stabilized using feedback on capacitive actuators. 
Electric shield (45 $\mu$m beryllium copper membrane) is set between test and source near mass.
Source mass vibrate horizontally.
 }
 \label{HUST}
\end{figure}

\begin{figure}[H]
 \begin{center}
  \includegraphics[width=70mm]{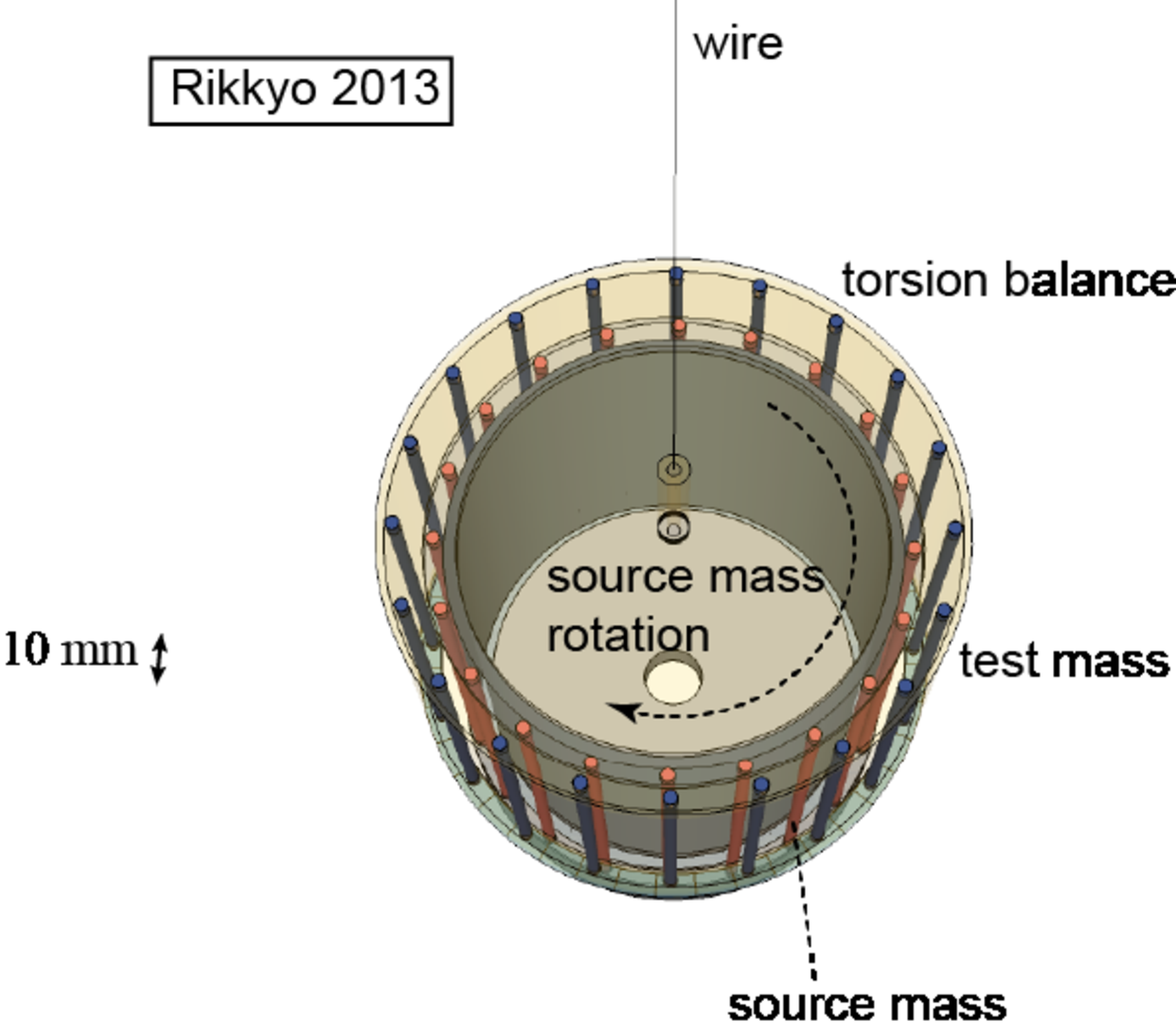}
 \end{center}
 \caption{Experiment of Rikkyo University group \cite{HE_News.32.2014}. 
 Test mass : 20 of 2 mm diameter, 55 mm long tungsten cylinder. 
 Source masses : 20 of 2 mm diameter and 70 mm long tungsten cylinder at $r =$ 5 - 9 mm (target center to source center distance). 
Torsion balance : 80 mm (target center-to-center) diameter cylinder. 
Wire : tungsten $40 \;\mu$m diameter, 236 mm long. 
Sensor : digital video imaging displacement sensor. Electric shield (400 $\mu$m aluminum pipe) is set between test and source masses.
Source mass rotates around the vertical axis.
 }
 \label{N4h}
\end{figure}

The group at Huazhong University of Science and Technology (HUST) performed torsion balance experiments (Figure \ref{HUST}) using parallel plates in null configurations
\cite{PhysRevLett.98.201101,PhysRevLett.108.081101}.
They obtained the best constraint of $\alpha < 1\times 10^{-3}$ at $\lambda=$ 1 mm.
The Rikkyo University group also performed torsion balance experiments (Figure \ref{N4h}) using a digital image sensor and obtained a preliminary result of $\alpha< 0.1$ at  $\lambda=$ 5 mm
\cite{HE_News.32.2014}.
This group is also testing the universality of free fall with a composition-dependent test
\cite{1742-6596-453-1-012007}.
\vspace{12pt}

\begin{figure}[H]
 \begin{center}
  \includegraphics[width=120mm]{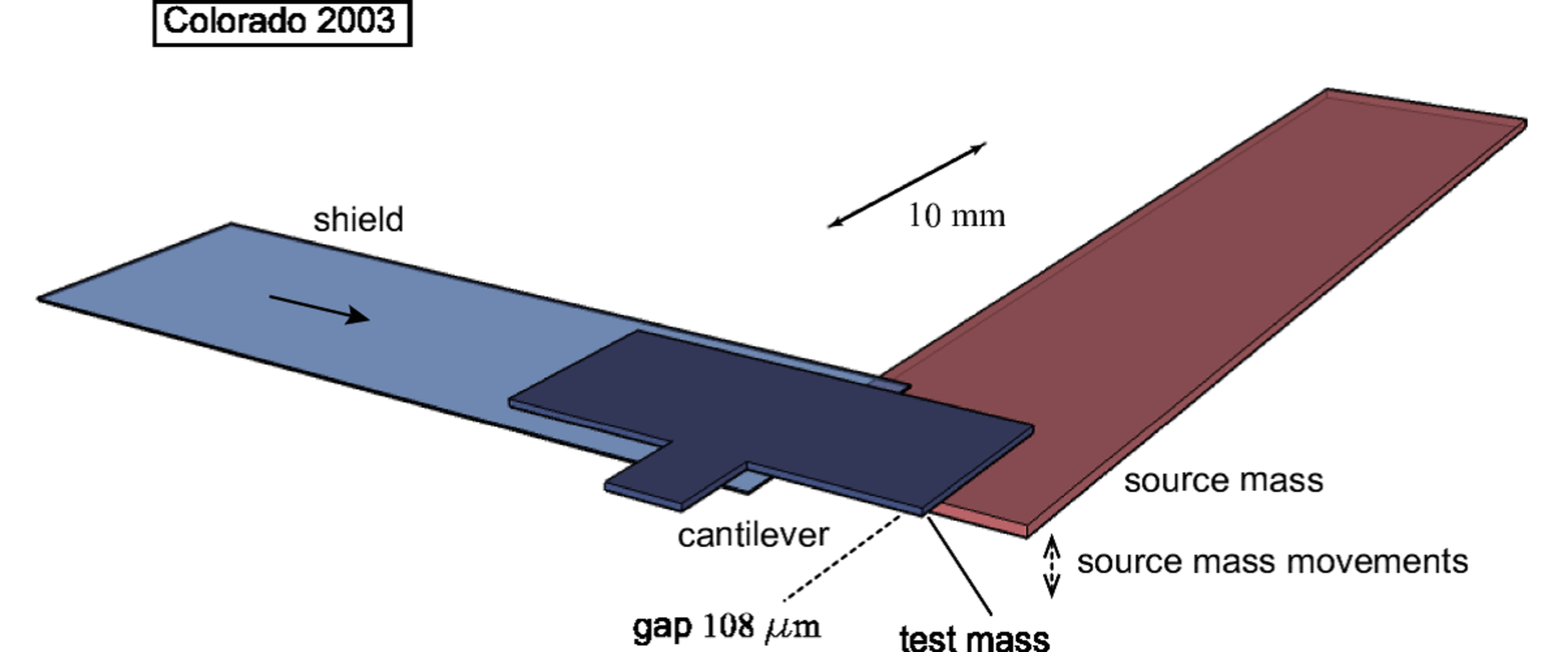}
 \end{center}
 \caption{Experiment of University of Colorado group \cite{Nature.01432}. 
 Test mass (tungsten plate): 5.1 mm long, 11.5 mm wide and 195 $\mu$m thick. 
Source masses (tungsten plate): 35 mm long, 7 mm wide and 305 $\mu$m thick, placed at  $d= 108 \;\mu$m (surface to surface) gap. Sensor : capacitive transducer. 
Electric shield (60 $\mu$m gold coated sapphire plate) is set between test and source masses.
Source mass vibrates vertically.
 }
 \label{Colorado}
\end{figure}

The University of Colorado group responded to the large extra-dimension model very quickly and obtained a result of $\alpha < 7$ at $\lambda=100 \;\mu\rm{m}$ (surface-to-surface separation) using a planar oscillator (Figure \ref{Colorado}) as a gravity sensor \cite{Long199923,Nature.01432}.
Unlike the HUST and Washington experiments, this experiment did not observe a nonzero gravity signal because of its sensitivity; therefore, only an upper limit on the gravitational force was determined. Similarly, all experiments that followed were performed at below 0.1 mm scales and could not observe a nonzero gravity signal.

\begin{figure}[H]
 \begin{center}
  \includegraphics[width=120mm]{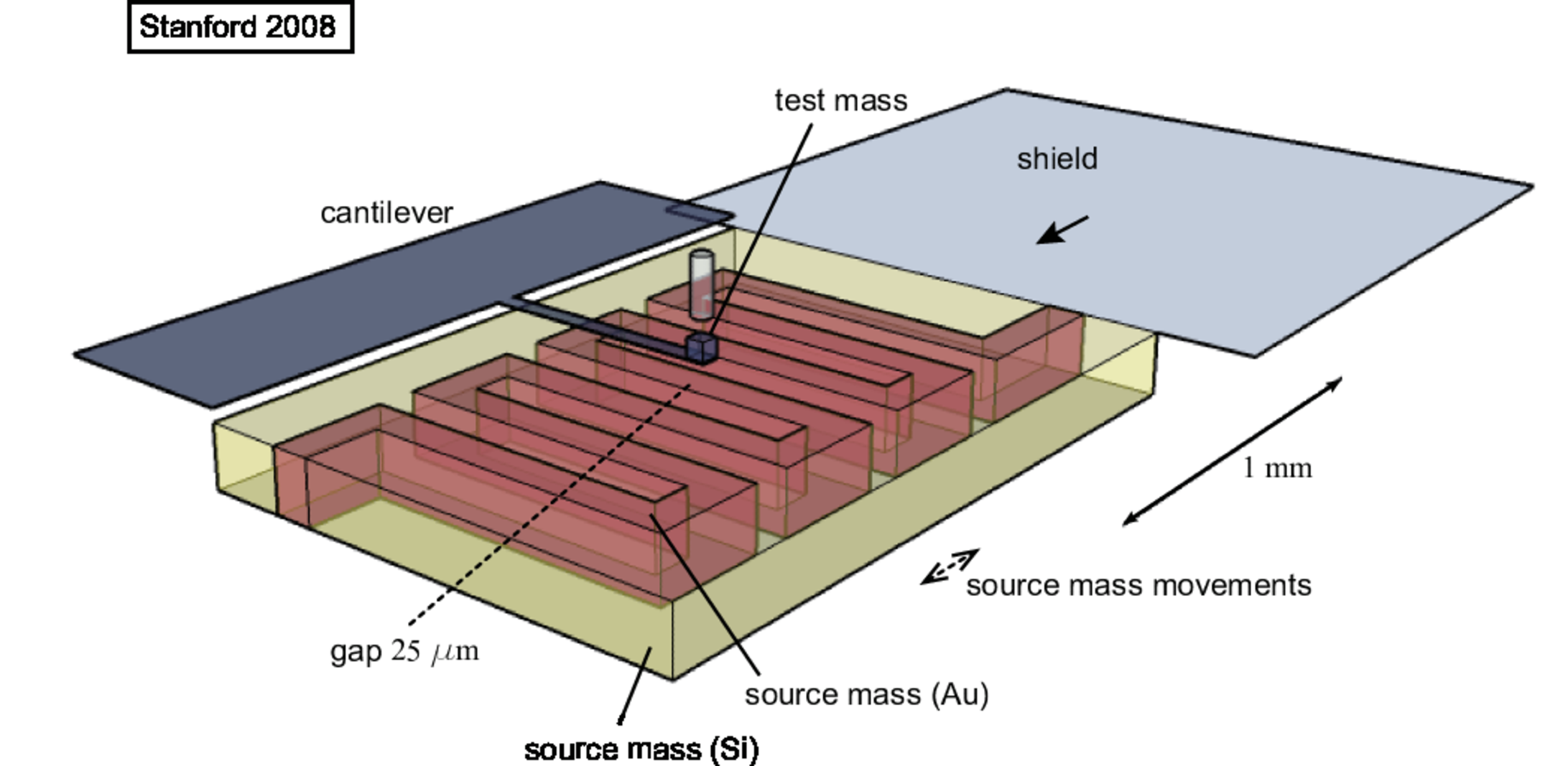}
 \end{center}
 \caption{Experiment of Stanford University group \cite{PhysRevLett.90.151101,PhysRevD.72.122001,PhysRevD.78.022002}. 
 Test mass (gold rectangular): 1.5 $\mu$g, $54 \times 54 \times 27 \;\mu m^3$. 
Source masses (gold and silicon bars): 1 mm long, 0.1 mm wide and 0.1 mm thick, placed at $d= 25 \mu$m (surface to surface) gap. Cantilever : silicon 250 $\mu$m long, 50 $\mu$m wide and 0.3 $\mu$m thick. Sensor : Fabry-Perot interferometer.
Electric shield (gold coated 3 $\mu$m silicon nitride membrane) is set between test and source masses.
Source mass moves reciprocating horizontally. 
}
 \label{Stanford}
\end{figure}

Very high-precision experiments were performed (Figure \ref{Stanford}) by the Stanford University group using cantilevers
\cite{PhysRevD.72.122001,PhysRevD.78.022002,PhysRevLett.90.151101}.
They obtained $\alpha < 500$ at $\lambda=25\; \mu$m (surface-to-surface separation).
Most recent experiments testing the large extra-dimension model achieved good precision at the separation gap $d$ because of the finite size effects shown in Figure \ref{Yukawa-model-plate}.

\subsection{Casimir force experiments}
In this section, we introduce experiments that provided the data in Figure \ref{alpha-lambda-microscale}.
Unlike torsion balance experiments in which $\alpha<1$ is obtained, the experimental sensitivities in this region are far beyond the strength of Newtonian gravity.
Therefore, all constraints were obtained as upper limits on the strength of the gravitational force, which corresponds to the experimental and theoretical precision $\sigma(F_C)$ for the dominant electric force $F_C$.
In the scale below $\mu$m, it is difficult to place electric shield between the source and test masses.
Upper limits on gravity can be estimated by
\begin{equation}
F_G = G(r)\frac{Mm}{r^2} \leq ( F_C^{exp} - F_C^{theory} ) \sim \sigma(F_C),
\label{estimation}
\end{equation}
provided the experimental results are consistent with the electric force calculation.
Then,
\begin{equation}
\gamma = \frac{G(r)}{G_N} \leq \frac{\sigma(F_C)}{F_{N}} =\left( \frac{F_C}{F_{N}} \right) \times \left( \frac{\sigma(F_C)}{F_C} \right)
\label{gamma-Casimir}
\end{equation}
can be estimated using the ratio of electric force to Newtonian gravity $F_C/F_{N}$ and a relative resolution $ \sigma(F_C)/F_C$ of the electric force; here, $\sigma(F_C)$ includes both experimental and theoretical errors.

\begin{figure}[t]
 \begin{center}
  \includegraphics[width=80mm]{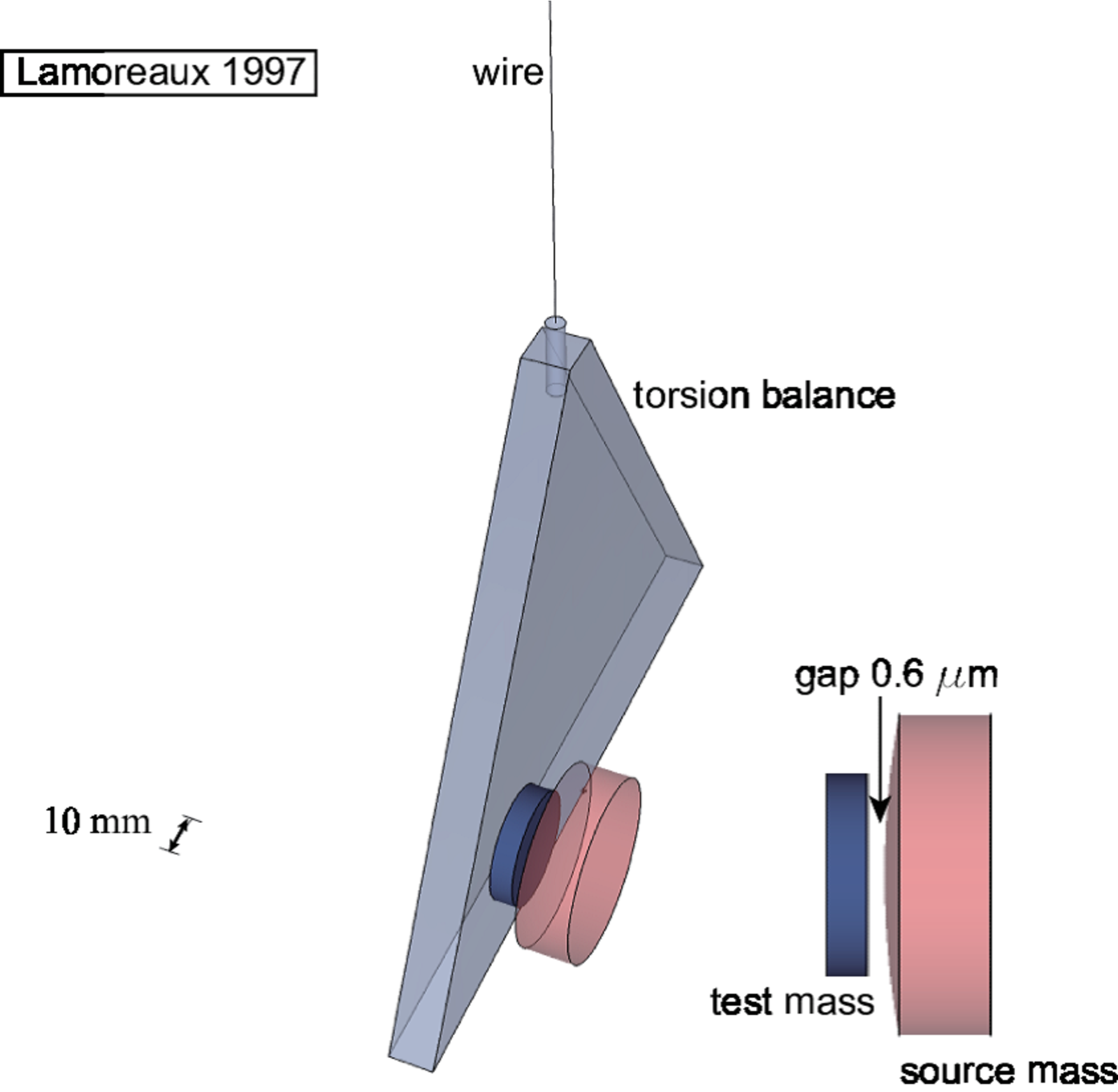}
 \end{center}
 \caption{ Experiment of Lamoreaux at University of Washington \cite{PhysRevLett.78.5}. Test mass : 2.5 cm diameter, 0.5 cm thick quartz optical flat (copper and gold coated). Source masses : 11.3 cm radius of curvature, 4 cm diameter  spherical lens (copper and gold coated). $d =$ 0.5 - 10 $\mu$m (surface to surface) gap. No shielding is applied between test and source mass. Wire : 76 $\mu$m, 66 cm tungsten wire. The torsion balance is kept static using feedback system with compensator plates. This experiment was not designed to test gravity. 
 }
 \label{Lamoreaux}
\end{figure}

\begin{figure}[b]
 \begin{center}
  \includegraphics[width=70mm]{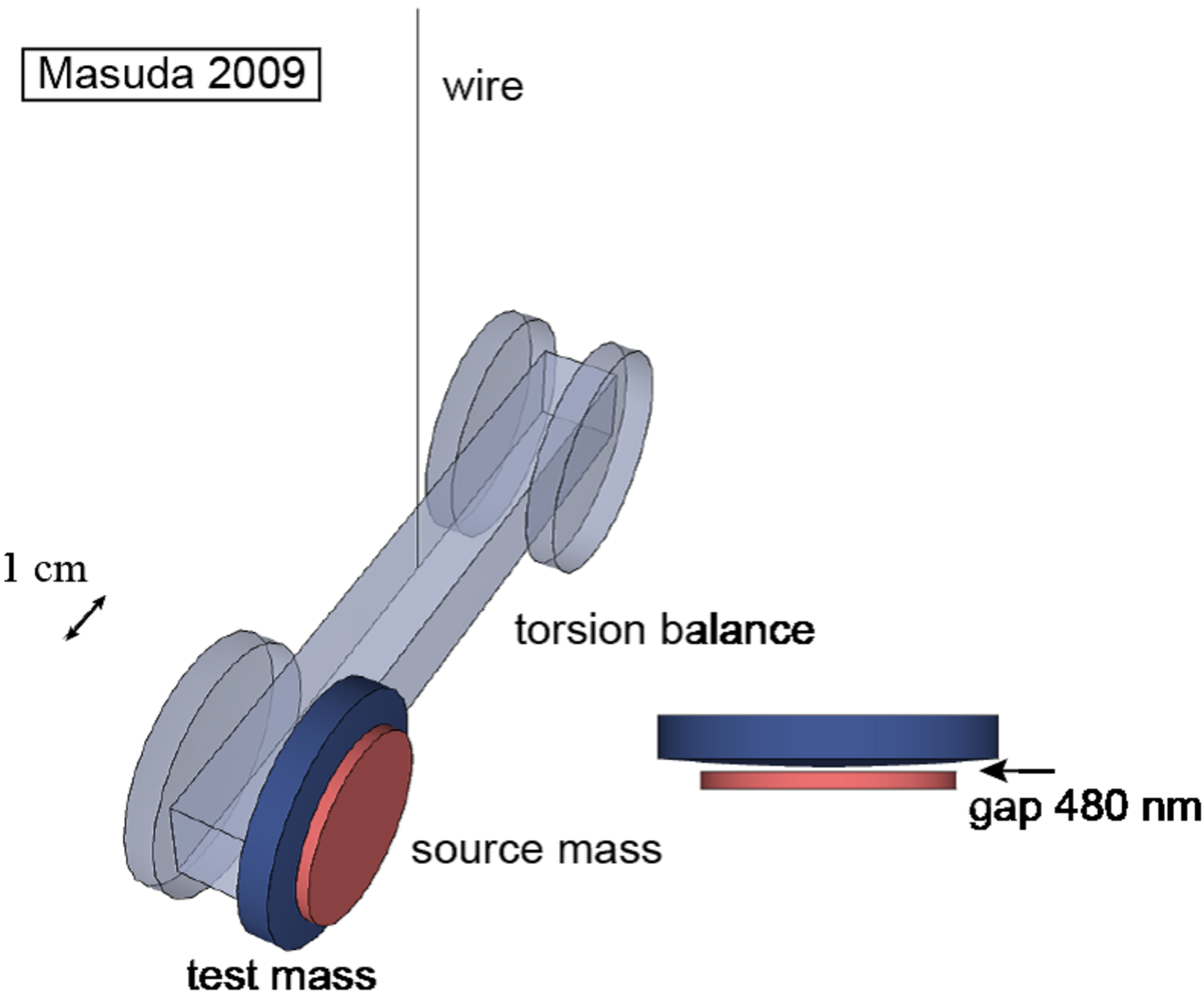}
 \end{center}
 \caption{Experiment of Masuda and Sasaki at ICRR, University of Tokyo \cite{PhysRevLett.102.171101}. Test mass : 20.7 cm radius of curvature, 2 cm diameter and 5 mm thick spherical lens. Source masses : 1.5 cm diameter, 2 mm thick optical flat. Test and source masses are 20 nm chromium and 1 $\mu$m gold coated. Surface to surface separation 0.5 - 7 $\mu$m. No shielding is applied between test and source mass. Wire : 60 $\mu$m diameter, 40 cm long tungsten wire. The torsion balance is kept static using feedback system. This experiment was designed to test gravity. }
 \label{Masuda}
\end{figure}

\begin{figure}[t]
 \begin{center}
  \includegraphics[width=80mm]{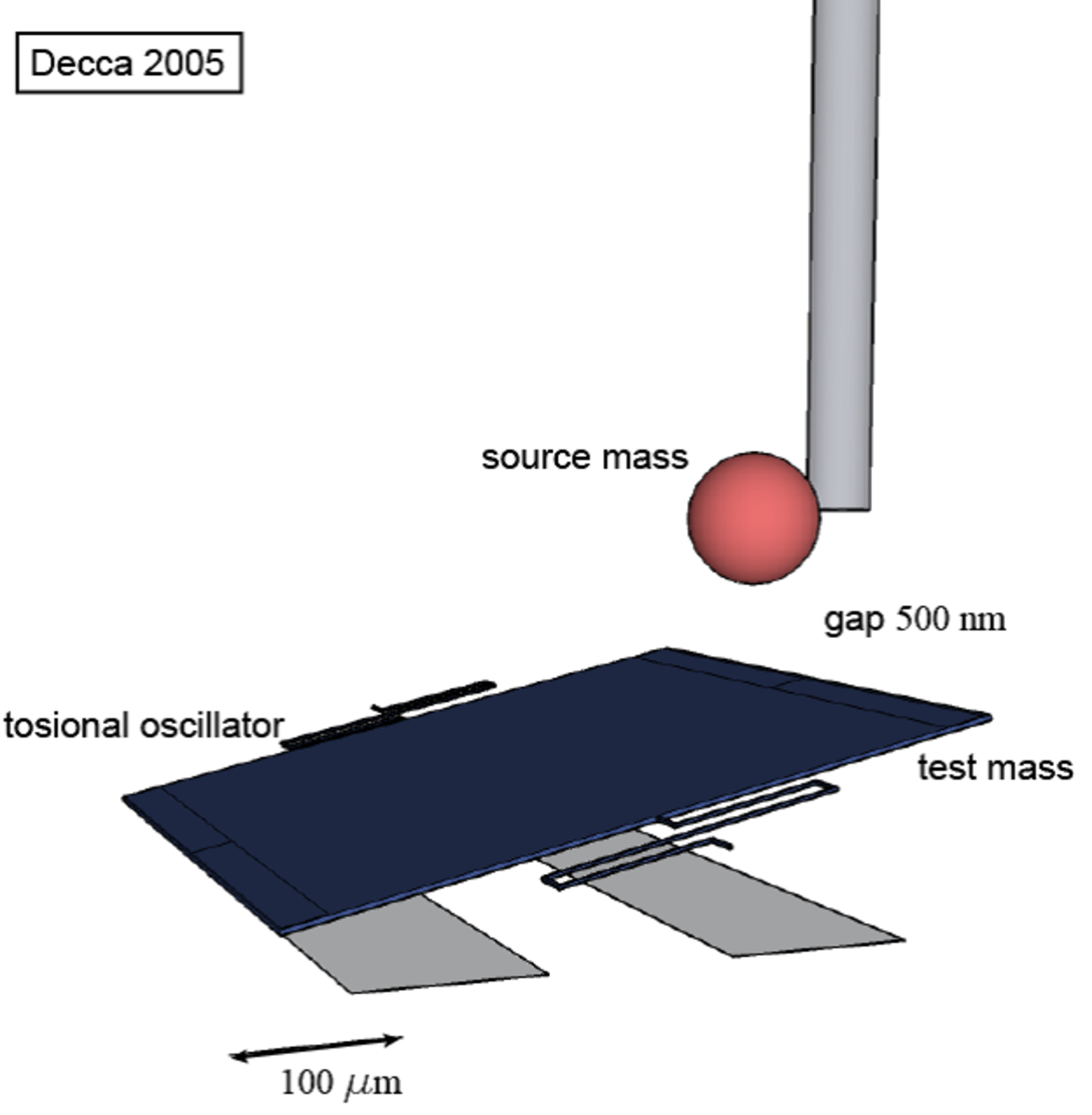}
 \end{center}
 \caption{Experiment of Decca et. al. \cite{PhysRevLett.94.240401}. Test mass : 200 nm thick gold or germanium layer with common 150 nm thick gold layer of microelectromechanical torsional oscillator. 
 Source masses : 50 $\mu$m radius sapphire sphere (chromium and gold coated) at $d = $ 150 - 500 nm (surface to surface) gap. No shielding is applied between test and source mass. This experiment is designed to cancel Casimir force ambiguity by measuring force difference between germanium and gold to search strong gravity. This experiment can also test composition dependence.
 }
 \label{Decca}
\end{figure}

\begin{figure}[h]
 \begin{center}
\vspace{1cm}
  \includegraphics[width=80mm]{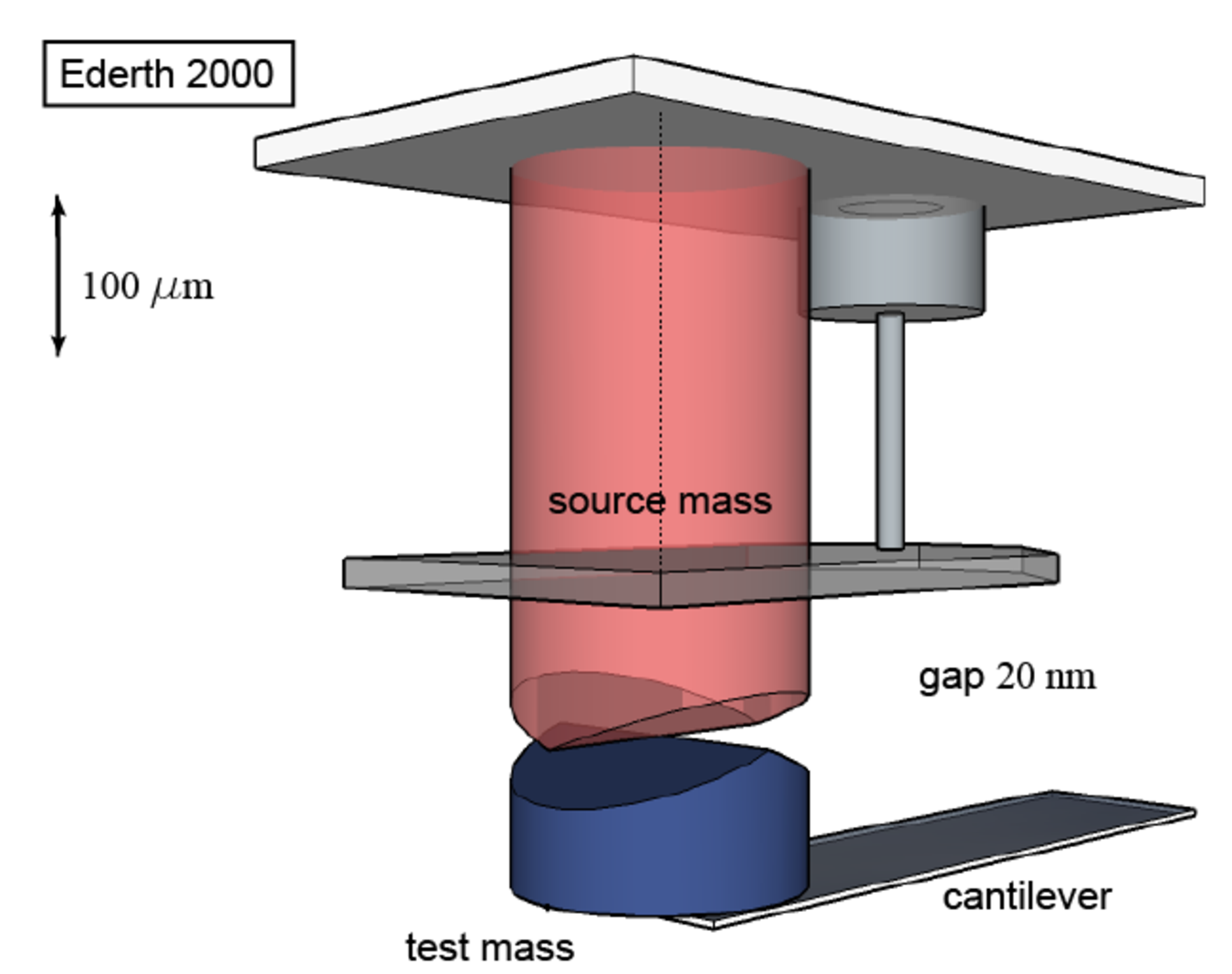}
 \end{center}
 \caption{Experiment of Ederth \cite{PhysRevA.62.062104}. Test mass and source mass are 10 mm radius cylindrical silica disks, on which 200 nm thick gold coated 10 - 15 $\mu$m mica sheets are glued. $d =$ 20 - 100 nm (surface to surface) gap. Cantilever : Bimorph spring. }
 \label{Ederth}
\end{figure}

There have been many Casimir force experiments that measure electric forces between two metal surfaces.
Because of the difficulty of preparing plates as thin as the measuring separation gap $d$, the data show the $\alpha-\lambda$ curve with kinks at $\lambda\sim d$, which are similar to that shown in Figure \ref{Yukawa-model-plate}.
Although not all experiments were intended to test the gravitational law, their results can be used to set the upper limits on $\alpha$ using (\ref{gamma-Casimir}).
Lamoreaux at the University of Washington performed Casimir force experiments (Figure \ref{Lamoreaux}) with  $\alpha<2 \times 10^{8}$ at $\lambda=5 \;\mu \rm{m}$ \cite{PhysRevLett.78.5} 
and later with $\alpha<2 \times 10^9$ at $\lambda=1 \;\mu \rm{m}$ 
\cite{PhysRevLett.107.171101}.
Masuda obtained (Figure \ref{Masuda}) $\alpha<3 \times 10^{10}$ at $\lambda=1 \;\mu \rm{m}$
\cite{PhysRevLett.102.171101}. 
Obrecht et al. also tested in the $\mu$m range, measuring Casimir-Polder forces
\cite{PhysRevLett.98.063201}, which were analyzed in \cite{PhysRevD.83.075004}. 
There are many other analyses by Bordaq and Mostepanenko et al.\cite{PhysRevD.47.2882,PhysRevD.58.075003,PhysRevD.62.011701,PhysRevD.63.115003,doi:10.1142/S0217751X02013356,PhysRevD.81.055003,PhysRevD.83.075004}.
Bao et al. also tested below the 200 nm range
\cite{PhysRevLett.105.250402}. 
Decca et. al. at Indiana-Purdue University tested 
 (Figure \ref{Decca}) with $\alpha<10^{13}$ at $\lambda=100 \;\rm{nm}$
\cite{PhysRevLett.94.240401,
PhysRevD.75.077101}. 
The University of California, Riverside group tested with $\alpha<10^{19}$ at $\lambda=10 \;\rm{nm}$
\cite{PhysRevA.62.052109}. 
Ederth in the Sweden Royal Institute of Technology tested  (Figure \ref{Ederth}) with $\alpha<5\times 10^{17}$ at $\lambda=20  \;\rm{nm}$
\cite{PhysRevA.62.062104}, 
which was analyzed in \cite{PhysRevD.63.115003}.
In addition to Casimir force measurements, van der Waals force measurements were performed \cite{Israelachvili21111972}, which were analyzed in \cite{Bordag199435}. 
These data approach $\alpha \sim 10^{30}$ at $\lambda \sim $ 1 nm.
As we saw in Figure \ref{alpha-lambda-explain}, these data do not strongly constrain the ADD model at the present precision, compared with the mm scale and collider experiments.


\subsection{Atomic, nuclear and particle experiments}
\label{subsec-atomic}
Atomic and nuclear systems can also be used to estimate the maximum allowed strength of gravity at microscopic scales using (\ref{gamma-Casimir}), where $F_C$ now represents the known standard model interactions.
For atomic systems, these interactions correspond to the Coulomb interaction between nuclei and electron (or other such negatively  charged particles for exotic atoms), and for nuclear systems, they correspond to the nuclear force between nucleons inside nuclei.
For an example, in an atomic system,
\begin{equation}
\gamma=\frac{G(r)}{G_N}<\left(\frac{\alpha_F \hbar c \frac{Z}{r^2}}
{G_N \frac{Mm}{r^2}}\right)\times
\left(\frac{\sigma(F_C)}{F_C}\right)
\end{equation}
can be obtained for a system in which a particle with mass $m$ and charge $-e$ is orbiting a nuclei with mass $M$ and charge $+Ze$.

Figure \ref{alpha-lambda-microscale} shows estimated results for hydrogen, the antiprotonic (pbar) helium atom, and the muonic hydrogen atom \cite{INPC2013}.
Their relative precisions
\begin{equation}
\left(
\frac{\sigma(F_C)}{F_C}
\right)^2
=
\left(
\frac{\sigma^{exp}(F_C)}{F^{exp}_C}
\right)^2
+
\left(
\frac{\sigma^{theory}(F_C)}{F^{theory}_C}
\right)^2
\end{equation}
can be estimated using relative errors in the experiments (e.g., error of atomic transition frequency) and theories, including physical constants (e.g., Planck constant, fine structure constant, and Rydberg constant).
These estimates were performed for the present paper using data from CODATA \cite{RevModPhys.84.1527} for the hydrogen atom, the ASACUSA experiment \cite{PhysRevLett.91.123401,PhysRevLett.96.243401} for the antiprotonic helium atom, and the PSI experiment \cite{Nature466} for the muonic hydrogen atom. The PSI data are not consistent with the known proton radius; nevertheless, the data can be used to set the gravitational upper limit. 
Details of the calculations will be published separately \cite{Tanaka-future}.
Several similar theoretical attempts have been made to constrain strong gravity at this scale for the muonic hydrogen atom \cite{doi:10.1142/S0218271814500059} and for the antiprotonic helium atom \cite{Salumbides201465,Bordag199435}.
Nuclear charge radii have also been used as a gravity sensor, in which a modification of the nuclear charge radius is estimated from strong gravity \cite{0954-3899-40-3-035107}.
An electron nuclear scattering experiment (the MTV-G experiment);  \cite{1742-6596-453-1-012018} was performed to constrain a strong geodetic precession of a scattering electron utilizing a detector setup from the MTV experiment for testing time reversal symmetry \cite{hyperfine}. 
Of these, the best precision was obtained from the antiprotonic helium spectroscopy data because of the heavy mass of the
antiproton. Constraints from a hydrogen system are limited, not by the experimental precision of $10^{-14}$ but by the precision for the Planck constant at $10^{-8}$.

\begin{figure}[h]
 \begin{center}
  \includegraphics[width=120mm]{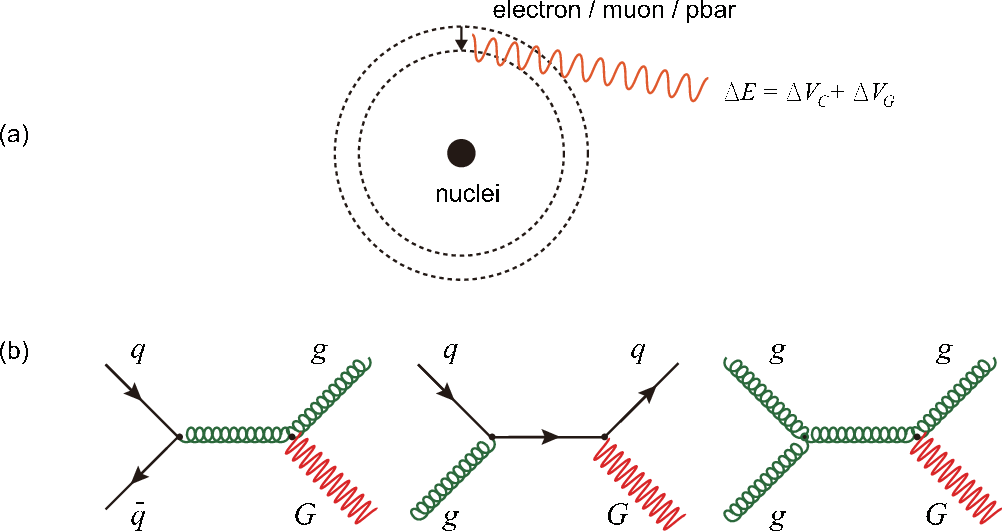}
 \end{center}
 \caption{Subatomic gravity sensors. (a) atomic systems : potential difference due to strong gravity is probed by transition energies. (b) collider experiments : monojet production channels $q\bar{q} \rightarrow gG, \;  qg \rightarrow qG, \; gg \rightarrow gG$ are searched. }
 \label{atomic}
\end{figure}

Similar to the atomic and nuclear tests, particle collision data can also be used to perform similar analyses that constrain the maximum allowed probability (cross section) of gravitational phenomena above standard model predictions.
Several graviton-producing channels can be used; however, the simplest is the real graviton emission $q\bar{q} \rightarrow gG, \;  qg \rightarrow qG$, and $gg \rightarrow gG$ shown in Figure \ref{atomic}.
The cross-sectional excess on a monojet produced from the standard model predictions can set upper limits on the gravitational coupling strength.
Here, the graviton ($G$) cannot be observed, unlike quarks ($q$) or gluons ($g$), creating QCD jets in the final states. 
In this channel, only one visible jet is observed; however, the emitted graviton is not observed, so the event is recognized by observing only one jet with large missing momentum.
By comparing with the standard model backgrounds, an upper limit on the monojet production cross section can be estimated.
Together with mini-black-hole production, which can be expected in strong gravity, results have  already been published for the LEP ($\sqrt{s}=189 - 209$ GeV);  \cite{arXiv:hep-ex0410004}, for the TEVATRON ($\sqrt{s}=1.96$ TeV);  \cite{PhysRevLett.101.181602,PhysRevLett.86.1156}, and for the LHC ($\sqrt{s}=7 - 8$ TeV); \cite{PhysRevLett.110.011802,Aad2011294,CMS-JHEP}.
In collider experiments, $M_D$ is the prime quantity to be determined as a function of $n$; then, $M_D$ is transformed to $\lambda$ using the ADD model (\ref{ADD-lambda-MD}).
All atomic, nuclear, and particle experiments cannot completely shield the standard model contributions (backgrounds); therefore, the resulting precision is poor because $F_C$ is much larger than $F_N$.
However, because of their extreme short measuring distances $r$, they can be sensitive to large values of $\lambda$ in power-law type models, as shown in Figure \ref{n-lambda}, which amplify $(\lambda/r)^n$ at a short range.

Among all microscopic experiments, the collider experiments are the only sensitive tool that can compete with the 0.1 mm scale experiments for $n=2$. For $n\geq 3$, none of the laboratory-scale experiments can compete with the collider experiments.

\section{Summary}
We have presented rough estimation procedures based on the Yukawa and power-law parametrizations.
Although the $\alpha-\lambda$ parametrization has been widely used because of its historical background, it is not always the best way to compare experimental constraints, especially for the large extra-dimension model.
A quantitative treatment based on the power-law parametrization has allowed us, for the first time, to compare results from collider experiments with those from short-range gravity experiments. The comparisons show that at $n=2$, stronger constraints can be obtained from 0.1 mm scale measurements than from collider experiments.

\section*{Acknowledgement}
Most of the art works were produced with the help of M. Hatori, S. Saiba, T. Sakuta, and E. Watanabe. Presented results from Rikkyo group were obtained by S. Ozaki, Y. Sakamoto, T. Yoshida, R. Tanuma, and Y. Totsuka for the MTV-G experiment, and by H. Murakami, K. Ninomiya for the short-range experiment. This work was supported by JSPS KAKENHI Grant-in-Aid for Challenging Exploratory Research, Grant Number 24654070.
The author would like to thank the editorial board of CQG for giving us the opportunity to write this review article.

\section*{References}
\bibliography{gravity}

\end{document}